\documentclass[12pt]{article}
\usepackage{graphicx}

\makeatletter
\def\lesssim{\mathrel{\mathpalette\vereq<}}
\def\gtrsim{\mathrel{\mathpalette\vereq>}}
\def\vereq#1#2{\lower3pt\vbox{\baselineskip1.5pt \lineskip1.5pt
\ialign{$\m@th#1\hfill##\hfil$\crcr#2\crcr\sim\crcr}}}
\makeatother

\title{Transport theory for interacting electrons  
connected to reservoirs}
 
\author{Akira Oguri
\\
 {\normalsize \em 
 Department of Material Science, 
Osaka City University,} \\
 {\normalsize \em  Sumiyoshi-ku, Osaka 558-8585, Japan 
 }
}

\begin{document}


\maketitle

\begin{abstract}
We describe microscopic theory for the 
quantum transport through finite interacting systems connected 
to noninteracting leads. It can be applied to small systems 
such as quantum dots, quantum wires, atomic chain, molecule, and so forth.  
The Keldysh formalism is introduced 
to study the nonlinear current-voltage characteristics,
and general properties of the nonequilibrium Green's functions are provided.
We apply the formulated to an out-of-equilibrium 
Anderson model that has  been used widely for 
the Kondo effect in quantum dots.
In the linear-response regime 
the Kubo formalism still has an advantage, 
because it relates the transport coefficients 
directly to the correlation functions defined with respect 
to the thermal equilibrium, 
and it has no ambiguities about a profile of the inner electric field.
We introduce a many-body transmission coefficient,  
by which the dc conductance can be expressed 
in a Landauer-type form, quite generally.
We also discuss transport properties of 
the Tomonaga-Luttinger liquid, 
which has also been an active field of research in this decade.

\end{abstract}

\newpage

\tableofcontents

\newpage

\section{Introduction}
\label{sec:Introduction}

Quantum transport through finite interacting-electron systems 
has been studied extensively in this decade. 
For instance, the Coulomb blockade and various 
effects named after Kondo, Aharanov-Bohm, Fano, Josephson, etc.\
in quantum dot systems have been a very active field of research.
Furthermore, the realization of 
non-Fermi liquid systems such as 
the Tomonaga-Luttinger liquid in quantum wires  
and multi-channel Kondo behavior in some novel systems 
have also been investigated by a number theorists.

To study transport properties of correlated electron systems,
theoretical approaches that can treat correctly 
both the interaction and quantum interference effects 
are required. 
The Keldysh Green's function approach is one of such methods 
\cite{Keldysh,Schwinger,KadanoffBaym,Caroli,Landau,Mahan,ChouSuHaoYu,Datta,Jauho}.
Specifically, the formulation for the nonlinear current-voltage 
profile by Caroli {\em et al\/} has been applied widely 
to the quantum transport phenomena.
In this report, we describe the outline of the 	Keldysh 
formalism in Sec.\ \ref{sec:Keldysh},
and then apply it to a single Anderson impurity, which is 
a standard model of quantum dots in the Kondo regime 
in Sec.\ \ref{sec:Anderson_model}.

When a finite sample is connected to reservoirs 
that can be approximated by free-electron systems 
with continuous energy spectrums, 
the low-energy eigenstates of whole the system 
including the attached reservoirs are determined coherently.
Thus, to understand the low-temperature properties, 
the information about the low-lying energy states 
of the whole system is required. 
The local Fermi-liquid theory \cite{Nozieres,YY},  
which was originally introduced for the Kondo systems \cite{Hewson},
is also applicable to the transport properties 
in wide classes of the interacting-electron systems
at low temperatures. 
In Sec.\ \ref{sec:Kubo},  
we reformulate the transport theory 
for the interacting systems connected to noninteracting leads 
based on the Kubo formalism. 
The dc conductance can be written in a Landauer-type form 
with a many-body transmission coefficient 
determined by a three-point correlation function. 
We also provide a brief introduction to 
Tomonaga-Luttinger model in Sec.\ \ref{sec:Tomonaga-Luttinger} 
to take a quick look at the transport properties 
of a typical interacting-electron system in one dimension.

\begin{figure}[tb]
\leavevmode 
\begin{center}
\begin{minipage}{0.75\linewidth}
\includegraphics[width=\linewidth]{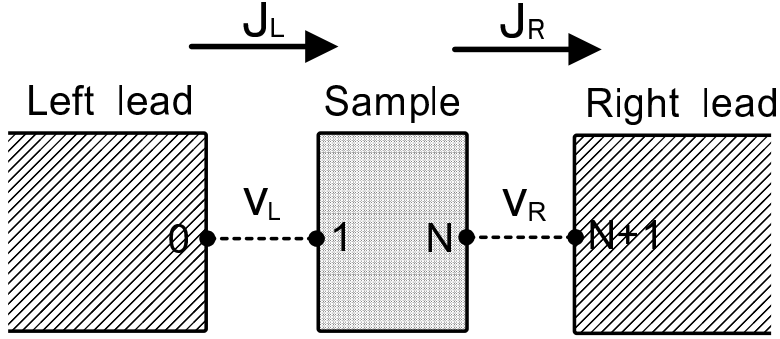}
\end{minipage}
\caption{Schematic picture of the system.} 
\label{fig:single}
\end{center}
\end{figure}

\section{Keldysh formalism for quantum transport} 
\label{sec:Keldysh}

\subsection{Thermal equilibrium}
\label{subsec:equilibrium}

We start with a system that consists of three regions: 
a finite central region ($C$) 
and two reservoirs on the left ($L$) and the right ($R$).
The central region consists of $N$ resonant levels, 
and the interaction  $U_{j_4 j_3; j_2 j_1}$ is 
switched on only for the electrons in this region.
We assume that each of the reservoirs is infinitely large 
and has a continuous energy spectrum.
The central region and reservoirs are connected with  
the mixing matrix elements $v_L^{\phantom{\dagger}}$ 
and $v_R^{\phantom{\dagger}}$, as illustrated in Fig.\ \ref{fig:single}. 
The complete Hamiltonian is given by 
\begin{eqnarray}
&&
\!\!\!\!\!\!\!\!\!\!
{\cal H}_{\rm tot}^{\rm eq}  
\,=\, {\cal H}_L \, + \, {\cal H}_R \, + \, {\cal H}_C^0 \,+ \, {\cal H}_C^U 
\, + \, {\cal H}_{\rm mix} \;. 
\label{eq:Heq}
\\
&&
\!\!\!\!\!\!\!\!\!\!
{\cal H}_L \,=\, -  \sum_{\scriptstyle ij\in L \atop \scriptstyle \sigma} 
        t_{ij}^L \,  c^{\dagger}_{i \sigma} c^{\phantom{\dagger}}_{j \sigma}
 \;, 
 \rule{0cm}{0.7cm}
 \qquad \quad
{\cal H}_R \, = \,  -  \sum_{\scriptstyle ij\in R \atop \scriptstyle \sigma} 
       t_{ij}^R \,  c^{\dagger}_{i \sigma} c^{\phantom{\dagger}}_{j \sigma}
      \;, \\
&& 
\!\!\!\!\!\!\!\!\!\!
 {\cal H}_C^{0} \,=  \, 
   -  \sum_{\scriptstyle ij\in C \atop \scriptstyle \sigma} 
       t_{ij}^C \,  c^{\dagger}_{i \sigma}\, c^{\phantom{\dagger}}_{j \sigma}
\rule{0cm}{0.6cm}
\;, \qquad 
   {\cal H}_C^{U} \, = \, 
   {1 \over 2} 
 \sum_{\scriptstyle \{j\} \in C \scriptstyle \atop \scriptstyle \sigma \sigma'}
    U_{j_4 j_3: j_2 j_1}\,  
 c^{\dagger}_{j_4 \sigma}\, c^{\dagger}_{j_3 \sigma'}
 c^{\phantom{\dagger}}_{j_2 \sigma'}\, c^{\phantom{\dagger}}_{j_1 \sigma}
\;, 
\label{eq:interaction}
\\
 && 
 \!\!\!\!\!\!\!\!\!\!
{\cal H}_{\rm mix} \,=\, 
-  \sum_{\sigma}\, v_L  \left[\,  
             c^{\dagger}_{0 \sigma} c^{\phantom{\dagger}}_{-1 \sigma}
             + \mbox{H.c.} \,\right] 
\, -  \sum_{\sigma} \, v_R  \left[\,  
             c^{\dagger}_{N+1 \sigma}\, c^{\phantom{\dagger}}_{N \sigma}
             + \mbox{H.c.} \,\right] 
  \;. 
  \rule{0cm}{0.5cm} 
\label{eq:tunnel}
\end{eqnarray}
Here, $c^{\dagger}_{j \sigma}$ 
 ($c^{\phantom{\dagger}}_{j \sigma}$) creates (destroys) 
an electron with spin $\sigma$ at site $j$,  
and $\mu$ is the chemical potential. Also, 
$t_{ij}^{L}$, $t_{ij}^{R}$, and $t_{ij}^{C}$ are 
the intra-region hopping matrix elements
in each of the three regions $L$, $R$, and $C$, respectively. 
The labels $1$, $2$, $\ldots$, $N$ are assigned to 
the sites in the central region.
Specifically, the label $1$ ($N$) is assigned to the site 
at the interface on the left (right), 
and the label $0$ ($N+1$) is assigned to the site 
at the reservoir-side of the left (right) interface.
We will be using units $\hbar=1$ unless otherwise noted.


The density matrix for 
the equilibrium state $\rho_{\rm eq}$ is given by 
\begin{eqnarray} 
\rho_{\rm eq} &=& 
e^{-\beta\{{\cal H}_{\rm tot}^{\rm eq} - \mu (N_L + N_C + N_R)\}} 
              \ / \   \mbox{Tr} \   
                 e^{-\beta\{{\cal H}_{\rm tot}^{\rm eq} - \mu (N_L + N_C + N_R)\}} 
\;, \rule{0cm}{0.6cm}
\label{eq:dens_eq}
\\
N_{L} &=&  \sum_{\scriptstyle i\in L, \sigma} 
c^{\dagger}_{i \sigma} c^{\phantom{\dagger}}_{i \sigma} \;, \quad
N_{C} \ = \  \sum_{\scriptstyle i\in C, \sigma} 
c^{\dagger}_{i \sigma}\, c^{\phantom{\dagger}}_{i \sigma} \;, \quad
N_{R} \ = \  \sum_{\scriptstyle i\in R, \sigma} 
c^{\dagger}_{i \sigma} c^{\phantom{\dagger}}_{i \sigma}\;.
\rule{0cm}{0.6cm}
\end{eqnarray}
Therefore, 
the Hamiltonian ${\cal H}_{\rm tot}^{\rm eq}$ and 
a single chemical potential $\mu$ determine the statistical weight
in thermal equilibrium. 

\setlength{\unitlength}{0.5mm}
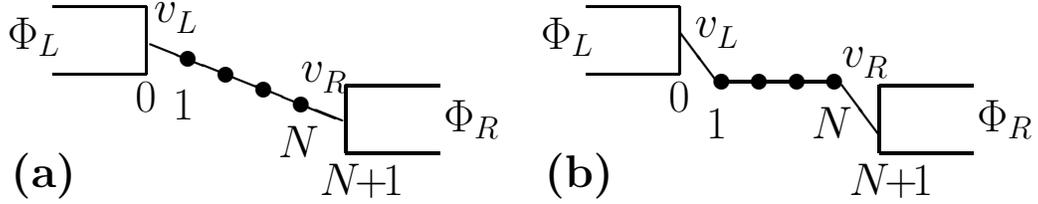
\begin{figure}[tb]
\ \hspace{-0.7cm}
\begin{minipage}[t]{8cm}
\begin{picture}(100,40)(-10,0)
\thicklines

\put(14,22){\line(1,0){25}}
\put(14,40){\line(1,0){25}}
\put(39,22){\line(0,1){18}}

\put(92,1){\line(1,0){25}}
\put(92,19){\line(1,0){25}}
\put(92,1){\line(0,1){18}}



\put(40,30){\line(5,-2){51}}

\put(50,26){\circle*{4}} 
\put(60,22){\circle*{4}} 
\put(70,18){\circle*{4}} 
\put(80,14){\circle*{4}} 

\put(2,27){\makebox(0,0)[bl]{\Large $\Phi_L$ }}
\put(118,5){\makebox(0,0)[bl]{\Large $\Phi_R$ }}

\put(41,33){\makebox(0,0)[bl]{\Large $v_L^{\phantom{1}}$ }}
\put(80,17){\makebox(0,0)[bl]{\Large $v_R^{\phantom{1}}$ }}

\put(36,12){\makebox(0,0)[bl]{\Large $0$}}
\put(46,10){\makebox(0,0)[bl]{\Large $1$}}
\put(74,0){\makebox(0,0)[bl]{\Large $N$}}
\put(85,-11){\makebox(0,0)[bl]{\Large $N\!\! + \!\! 1$}}
\put(3,-11){\makebox(0,0)[bl]{\Large \bf (a)}}
\end{picture}
\end{minipage}
\begin{minipage}[t]{8cm}

\begin{picture}(100,40)(10,0)
\thicklines

\put(14,22){\line(1,0){25}}
\put(14,40){\line(1,0){25}}
\put(39,22){\line(0,1){18}}


\put(92,1){\line(1,0){25}}
\put(92,19){\line(1,0){25}}
\put(92,1){\line(0,1){18}}

\put(39,33){\line(3,-4){10}}
\put(82,19){\line(3,-4){10}}

\put(50,20){\line(1,0){30}}

\put(50,20){\circle*{4}} 
\put(60,20){\circle*{4}} 
\put(70,20){\circle*{4}} 
\put(80,20){\circle*{4}} 

\put(2,27){\makebox(0,0)[bl]{\Large $\Phi_L$ }}
\put(118,5){\makebox(0,0)[bl]{\Large $\Phi_R$ }}

\put(43,29){\makebox(0,0)[bl]{\Large $v_L^{\phantom{1}}$ }}
\put(82,21){\makebox(0,0)[bl]{\Large $v_R^{\phantom{1}}$ }}

\put(36,12){\makebox(0,0)[bl]{\Large $0$}}
\put(46,5){\makebox(0,0)[bl]{\Large $1$}}
\put(74,5){\makebox(0,0)[bl]{\Large $N$}}
\put(84,-11){\makebox(0,0)[bl]{\Large $N\!\! + \!\! 1$}}
\put(3,-11){\makebox(0,0)[bl]{\Large \bf (b)}}

\end{picture}
\end{minipage}
\medskip

\caption{
Examples for the profile of the electrostatic potential $\Phi_C(i)$ for  
 (a) an insulating sample, and  (b) a metal sample, where ${\sl e} V = \Phi_L -  \Phi_R$.} 

\label{fig:inner_field}
\end{figure}

\subsection{Statistical weight for nonequilibrium steady states}

When the voltage $V$ is applied, 
the contribution of the electrostatic potential has 
to be included into ${\cal H}_{\rm tot}^{\rm eq}$, as 
\begin{eqnarray}
{\cal H}_{\rm tot}  
&=&  {\cal H}_{\rm tot}^{\rm eq} \, + \, {\cal V}_{\rm ext} \;,
\label{eq:H_neq}
\\
{\cal V}_{\rm ext} &=&  
        \Phi_L\,  N_L  \,  + \,   \Phi_R\,  N_R 
   \, + \,    \sum_{\scriptstyle i\in C, \sigma}  \Phi_{C}(i) \,
\;. 
\label{eq:H_under_bias}
\rule{0cm}{0.6cm}
\end{eqnarray}
Here 
$\Phi_L$ and $\Phi_R$ are the potentials 
for the lead at $L$ and $R$, respectively,    
and the applied bias voltage corresponds to 
${\sl e}V \equiv \Phi_L -  \Phi_R$. 
To determine the potential profile in the central region $\Phi_{C}(i)$,
the energy of the electric field should also be included into
the Hamiltonian eq.\ (\ref{fig:inner_field}),
 and it should be determined self-consistently. 
However for simplicity, we assume that $\Phi_{C}(i)$ 
is a given function.
In Fig.\ \ref{fig:inner_field}, two typical profiles are illustrated. 
 For an insulating sample,
 there must be a finite electric field in the central region 
 and the potential shows approximately a linear $i$-dependence 
 as that in the panel (a). 
 In an opposite case, for a metallic sample, 
 the electric field vanishes in the central region, 
  and the potential profile will become the one as shown in the panel (b). 
Realistic situations seem to be in between these two extreme cases.

In contrast to the thermal equilibrium,
one cannot write down the density matrix that
describes the nonequilibrium statistical weight simply 
by using a single chemical potential, 
\begin{equation} 
\rho \neq e^{-\beta\{{\cal H}_{\rm tot} - \mu (N_L + N_C + N_R)\}} 
              \ / \   \mbox{Tr} \   
                 e^{-\beta\{{\cal H}_{\rm tot} - \mu (N_L + N_C + N_R)\}} 
\;, 
\label{eq:dens_neq}
\end{equation} 
because this statistical weight describes the situation after the 
electrons have already been redistributed to gain 
the electrostatic potential energy. 
One possible statistical weight that describes 
a nonequilibrium steady state 
was introduced by Caroli {\em et al\/} \cite{Caroli}.


In the formulation of Caroli {\em et al\/},
the coupling to the leads ${\cal H}_{\rm mix}$ and 
the interaction ${\cal H}_C^U$ in the sample region 
are switched on adiabatically by separating the 
total Hamiltonian in the form
\begin{eqnarray}
{\cal H}_{\rm tot}(t)  
&=&  {\cal H}_1 \ + \ {\cal H}_2(t) 
\label{eq:H_neq_2} 
\;, \\ 
{\cal H}_1 
&=& {\cal H}_{1;L} + {\cal H}_{1;C} + {\cal H}_{1;R} 
\;,  \rule{0cm}{0.5cm}
\label{eq:H_0a} 
 \\
{\cal H}_{1;L} &=& {\cal H}_L + \Phi_L\,  N_L 
\;, \qquad \quad 
{\cal H}_{1;R} \ = \ {\cal H}_R  + \Phi_R\,  N_R 
\;,  \rule{0cm}{0.5cm}
\\
{\cal H}_{1;C} &=& {\cal H}_C^0 
   +     \sum_{\scriptstyle i\in C, \sigma} \Phi_{C}(i) \,
         c^{\dagger}_{i \sigma}\, c^{\phantom{\dagger}}_{i \sigma} 
\;,  \rule{0cm}{0.5cm} \\
{\cal H}_2(t) 
 &=&  \left[\,  {\cal H}_{\rm mix} + {\cal H}_C^U\, \right]\, e^{-\delta |t|}
\rule{0cm}{0.5cm}
\;.
\label{eq:H_0b} 
\end{eqnarray}
Here, $\delta=0^+$ is an positive infinitesimal.
Because ${\cal H}_1$ has a quadratic form,
it is possible to use the Wick theorem 
in the perturbation expansion with respect to ${\cal H}_2$. 
At $t=-\infty$ the two reservoirs and the impurity are isolated,
so that the different chemical potentials $\mu_{L}$, 
 $\mu_{R}$, and $\mu_{C}$ can be introduced 
 into the three regions in the initial condition.
The time evolution of the density matrix is determined by 
the equation 
\begin{equation} 
{\partial  \over \partial t} \, \rho(t) \ = \ -i \, 
\biggl[\, {\cal H}_{\rm tot}(t) \ , \ \rho(t) \, \biggr] 
\;.
\label{eq:Neumann}
\end{equation}
The formal solution of this equation 
can be obtained, by using the interaction 
representation $\widetilde{\rho}(t)   =   
        e^{i\,{\cal H}_1 t} \, \rho(t)\, e^{-i\,{\cal H}_1 t}$ 
and $\widetilde{{\cal H}}_2(t)  =  
      e^{i\,{\cal H}_1 t}\, {\cal H}_2(t)\, e^{-i\,{\cal H}_1 t}$, 
      as 
\begin{eqnarray} 
&&
\!\!\!\!\!\!\!\!\!
\widetilde{\rho}(t)  \,=\, U(t,t_0) \, \widetilde{\rho}(t_0) \, U(t_0,t) 
\label{eq:rho} \;,
\rule{0cm}{0.6cm}
\\
&&
\!\!\!\!\!\!\!\!\!
U(t,t_0) \,=\, \mbox{T} \exp 
        \left[\,- i \int_{t_0}^t dt' \ \widetilde{\cal H}_2 (t') \,\right] 
\;,
\rule{0cm}{0.7cm} 
\label{eq:U_2}
\\
&&
\!\!\!\!\!\!\!\!\!
U(t_0,t) \,=\,\widetilde{\mbox{T}} \exp 
        \left[\, i \int_{t_0}^t dt' \ \widetilde{\cal H}_2 (t') \,\right] 
\label{eq:U_2_tild}\;.
\rule{0cm}{0.7cm} 
\end{eqnarray}
Here, T denotes the operator for the chronological time order, 
and $\widetilde{\mbox{T}}$ is the anti-time-ordering operator.
Note that eq.\ (\ref{eq:U_2_tild}) is the Hermite conjugatae of
eq.\ (\ref{eq:U_2}). 
Caroli {\em et al} have assumed that the initial condition 
at $t_0 \to -\infty$ is given by
\begin{equation}
\widetilde{\rho}(-\infty)
\ =  \ 
  {
       e^{-\beta\{ {\cal H}_{1;L}- \mu_L N_L\}} 
    \  e^{-\beta\{ {\cal H}_{1;C}- \mu_C N_C\}} 
    \  e^{-\beta\{ {\cal H}_{1;R}- \mu_R N_R\}} 
  \over
     \mbox{Tr} \left[\,
       e^{-\beta\{ {\cal H}_{1;L}- \mu_L N_L\}} 
    \  e^{-\beta\{ {\cal H}_{1;C}- \mu_C N_C\}} 
    \  e^{-\beta\{ {\cal H}_{1;R}- \mu_R N_R\}} 
               \,\right]
  }
\;.
\label{eq:rho_ini}
\end{equation}
Namely, at $t_0 \to -\infty$, each system 
is in a thermal equilibrium state 
with the chemical potential $\mu_{L,R,C}$.
Here $\mu_L-\mu_R = \Phi_L-\Phi_R = {\sl e} V$.

\subsection{Perturbation expansion along the Keldysh contour}

\begin{figure}[tb]
\setlength{\unitlength}{1mm}
\begin{center}
\begin{picture}(120,20)
\thicklines

\put(100,10){\oval(12,9)[r]} 

\put(60,5.5){\vector(1,0){1}}
\put(58.5,14.5){\vector(-1,0){1}}

 \put(20,14.5){\line(1,0){80}}
 \put(20,5.5){\line(1,0){80}}

 \put(7,14){\makebox(0,0)[bl]{\large $-\infty$}}
 \put(7,4){\makebox(0,0)[bl]{\large $-\infty$}}
 \put(110,9){\makebox(0,0)[bl]{\large $+\infty$}}

 \put(66,1){\makebox(0,0)[bl]{$-$branch}}
 \put(66,17){\makebox(0,0)[bl]{$+$branch}}

\end{picture}
\end{center}

\caption{The Keldysh contour for the time evolution.}
\label{fig:loop}
\end{figure}
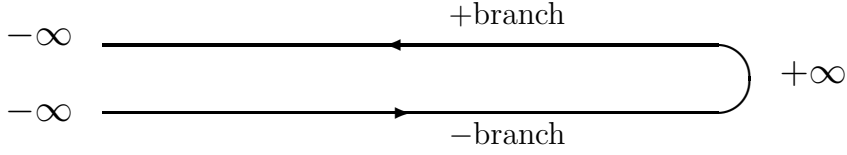

The perturbed part $\widetilde{\cal H}_2(t)$ is switched fully 
on at $t=0$.  
Therefore the expectation value of physical quantities are  
defined with respect to the density matrix at $t=0$, 
 \begin{eqnarray}
\langle {\cal O} \rangle &\equiv& 
            \mbox{Tr} \left[\,\rho(0) \,{\cal O}_S \, \right]  
\label{eq:average}
\nonumber \\    
     &=&  \mbox{Tr} \left[\,\widetilde{\rho}(0) \,{\cal O}_S \,\right]
      \  = \  \mbox{Tr} 
      \left[\,\widetilde{\rho}(-\infty) \, U(-\infty,0) \, 
            {\cal O}_S \, U(0,-\infty) \, \right]
\label{eq:O}
\;,
\end{eqnarray}
where ${\cal O}_S$ is a Schr\"{o}dinger operator, and
 $\left[\,\rho(0),\, {\cal H}_{\rm tot}(0) \, \right] =0$ 
for the stationary states.
Equation (\ref{eq:O}) can be rewritten by using a property 
of the time-evolution operator, 
 $U(-\infty,0) = U(-\infty,+\infty)\,U(+\infty,0)$, as 
\begin{eqnarray}
\langle {\cal O} \rangle \,=\, 
      \langle\, U(-\infty, +\infty) 
            \; U(+\infty,0)\, {\cal O}_S \, U(0,-\infty) \, \rangle_0
\label{eq:O2}
\;,
\end{eqnarray}
where $\langle \cdots \rangle_0 \equiv 
\mbox{Tr} \left[\,\widetilde{\rho}(-\infty) \cdots \right]$.
The stream of the time in eq.\ (\ref{eq:O2}) 
can be mapped onto a loop shown in Fig.\ \ref{fig:loop}:
starting from $t=-\infty$, observing a quantity
${\cal O}$ at $t=0$, then proceeding to $t=+\infty$, 
and then going  back to $t=-\infty$.
In the case of the usual $T=0$ Green's function with respect to  
the equilibrium ground state, 
the wavefunction at $t=+\infty$ 
is essentially the same with the one at $t=-\infty$ 
apart from a phase factor 
if the initial state has no degeneracy \cite{FetterWalecka}.
Thus, eq.\ (\ref{eq:O2}) can be decoupled at the time $t=+\infty$.
However, this simplification does not take place 
in the nonequilibrium case of the initial condition eq.\ (\ref{eq:rho_ini}),
Therefore, one has to treat the time loop 
including the way back to $t \to -\infty$.

The time-dependent expectation value is 
defined by using the Heisenberg operator  ${\cal O}_H(t)$, 
and is written in the form   
\begin{eqnarray}
\langle {\cal O}(t) \rangle &\equiv& 
            \mbox{Tr} \left[\,\rho(0) \,{\cal O}_H(t) \, \right]  
\label{eq:average_H}
\nonumber\\    
     &=&  
       \langle \, U(-\infty,+\infty) \, 
      U(+\infty,t) \, \widetilde{\cal O}(t) \, U(t,-\infty) \, \rangle_0
\nonumber\\
     &=&  
     \langle \, U(-\infty,+\infty) \, 
    \left\{\, \mbox{T}\ U(+\infty,-\infty) 
    \,\widetilde{\cal O}(t) \, \right\} \rangle_0
\label{eq:O_H} 
\nonumber\\
     &=&  
     \langle\, \mbox{T}_{\mbox{c}}\, 
      \, U_c \,\widetilde{\cal O}(t^-) \, \rangle_0
\label{eq:O_HC}
\;.
\end{eqnarray}
Here, the relation among the Schr\"{o}dinger  ${\cal O}_S$,
interaction $\widetilde{\cal O}(t)$,
and Heisenberg ${\cal O}_H(t)$ representations are given by 
\begin{equation}
\widetilde{\cal O}(t) \, = \, 
 e^{i\,{\cal H}_1t}\, {\cal O}_S \,e^{-i\,{\cal H}_1 t} \;,
  \qquad 
  {\cal O}_H(t) \,= \, U(0,t) \, \widetilde{\cal O}(t) \, U(t,0) \;.
\end{equation}
In eq.\ (\ref{eq:O_HC}), 
 $\mbox{T}_{\mbox{c}}$ and $U_c$ 
express the time order and time evolution 
along the Keldysh contour, respectively,
and $t^-$ denotes the time in the $-$branch in Fig.\ \ref{fig:loop}.

The perturbation expansion 
with respect to ${\cal H}_2$ can be carried out by  
substituting eqs.\ (\ref{eq:U_2}) and (\ref{eq:U_2_tild}), 
respectively, into $U(+\infty,-\infty)$ and $U(-\infty,+\infty)$
in eq.\ (\ref{eq:O_H}), as 
\begin{eqnarray}
&&
\!\!\!\!\!\!\!\!\!\!\!\!\!\!\!\!\!
\langle {\cal O}(t) \rangle \,=\, 
 \sum_{n=0}^{\infty}  \sum_{m=0}^{\infty}  
  \, {i^{n} \over n!}\ {(-i)^m \over m!} \,
  \int_{-\infty}^{+\infty} dt_1' \cdots dt_{n}' \, 
  \int_{-\infty}^{+\infty} dt_1 \cdots dt_{m} 
\nonumber \\
   & & \quad  \times \, 
    \langle \,  \left\{ \widetilde{\mbox{T}}\  
    \widetilde{\cal H}_2(t_1') \cdots  \widetilde{\cal H}_2(t_{n}') \right\} 
    \left\{ \mbox{T}\  
    \widetilde{\cal H}_2(t_1) \cdots  \widetilde{\cal H}_2(t_{m})  
     \,\widetilde{\cal O}(t) \right\} \, \rangle_0
\rule{0cm}{0.6cm}
\label{eq:O_H_expand} 
\;.
\end{eqnarray}
The Wick's theorem is applicable to 
the average $\langle \cdots \rangle_0$ 
because ${\cal H}_1$ has a bilinear form.
However, 
because $U(+\infty,-\infty)$ gives a factor $(-i)^m$ 
for the $m$-th order terms 
while $U(-\infty,+\infty)$ gives a factor $(+i)^n$ for the $n$-th order 
terms, four types of the Green's functions are necessary 
to distinguish the contributions from these two branches.

\subsection{Nonequilibrium Green's function}

We now introduce the four types of the Green's functions,
which are required in the Feynman-diagrammatic approach to
the perturbation expansion along the time loop,
\begin{eqnarray}
G^{--}_{ij} (t_1, t_2) &\equiv& -i\, 
\langle \,\mbox{T}\,
c^{\phantom{\dagger}}_{i \sigma}(t_1)\, c^{\dagger}_{j \sigma}(t_2)
\,\rangle 
\nonumber\\
&=& -i\,
      \langle\,
      \mbox{T}_{\mbox{c}}\, 
      \, U_c \, 
      \widetilde{c}^{\phantom{\dagger}}_{i \sigma}(t_1^-)\, 
      \widetilde{c}^{\dagger}_{j \sigma}(t_2^-)
      \, \rangle_0
\;,
\\
G^{++}_{ij} (t_1, t_2) &\equiv& -i\, 
\langle \,\widetilde{\mbox{T}}\,
c^{\phantom{\dagger}}_{i \sigma}(t_1)\, c^{\dagger}_{j \sigma}(t_2)
\,\rangle
\rule{0cm}{0.6cm}
\nonumber\\
&=& -i\,
      \langle\,
      \mbox{T}_{\mbox{c}}\, 
      \, U_c \, 
      \widetilde{c}^{\phantom{\dagger}}_{i \sigma}(t_1^+)\, 
      \widetilde{c}^{\dagger}_{j \sigma}(t_2^+)
      \, \rangle_0
\;,
\\
G^{+-}_{ij} (t_1, t_2) &\equiv& -i\, 
\langle 
c^{\phantom{\dagger}}_{i \sigma}(t_1)\, c^{\dagger}_{j \sigma}(t_2)
\,\rangle 
\rule{0cm}{0.6cm}
\nonumber\\
&=& -i\,
      \langle\,
      \mbox{T}_{\mbox{c}}\, 
      \, U_c \, 
      \widetilde{c}^{\phantom{\dagger}}_{i \sigma}(t_1^+)\, 
      \widetilde{c}^{\dagger}_{j \sigma}(t_2^-)
      \, \rangle_0
\;,
\\
G^{-+}_{ij} (t_1, t_2) &\equiv&  \phantom{-} i\, 
\langle 
c^{\dagger}_{j \sigma}(t_2) \, c^{\phantom{\dagger}}_{i \sigma}(t_1) 
\,\rangle 
\rule{0cm}{0.6cm}
\label{eq:G<}
\nonumber\\
&=& -i\,
      \langle\ 
      \mbox{T}_{\mbox{c}}\, 
      \, U_c \, 
      \widetilde{c}^{\phantom{\dagger}}_{i \sigma}(t_1^-)\, 
      \widetilde{c}^{\dagger}_{j \sigma}(t_2^+)
      \ \rangle_0
\;.
\end{eqnarray}
Here $c^{\phantom{\dagger}}_{i \sigma}(t_1)$ 
and $c^{\dagger}_{j \sigma}(t_2)$ are Heisenberg operators.
$t^{+,-}$ denotes the time $+$ or $-$branch in Fig.\ \ref{fig:loop}. 
For these Green's functions,
there is another notation used widely in literatures, and  
the relation between that and the present one by 
Lifshitz-Pitaevskii \cite{Landau} is 
summarized in Table 1.

\begin{table}[bt]
\begin{center}
\begin{tabular}{|l||c|c|c|c|} \hline 
Lifshitz-Pitaevskii \rule{0cm}{0.5cm} 
& $G^{--}$ & $G^{++}$ & $G^{+-}$ & $G^{-+}$ \\ 
\hline \hline
Alternative notation\rule{0cm}{0.5cm}
& $G_c$ & $\widetilde{G}_c$ & $G^{>}$ & $G^{<}$  \\ 
\hline
\end{tabular}
\caption{Correspondence between two standard notations.}
\end{center}
\end{table}

Based on the Feynman-diagrammatic approach, 
the Dyson equation can be expressed in a $2\times 2$ matrix form;
\begin{eqnarray}
    \mbox{\boldmath $G$}_{ij}(\omega) &=& 
    \mbox{\boldmath $g$}_{ij}(\omega) \ + \ \sum_{lm} \,  
    \mbox{\boldmath $g$}_{il}(\omega)\,  
    \mbox{\boldmath $\Sigma$}_{lm}(\omega) \,
    \mbox{\boldmath $G$}_{mj}(\omega)      \;, 
\label{eq:Dyson}
\\
 \mbox{\boldmath $G$}_{ij} &=& \left[ 
 \matrix { G_{ij}^{--} & G_{ij}^{-+}   \cr
           G_{ij}^{+-} & G_{ij}^{++}  \cr  }
                             \right]
\;,  \qquad
\mbox{\boldmath $\Sigma$}_{lm} \ = \ \left[ 
 \matrix { \Sigma_{lm}^{--} & \Sigma_{lm}^{-+}   \cr
           \Sigma_{lm}^{+-} & \Sigma_{lm}^{++}  \cr  }
                             \right] .
\rule{0cm}{0.8cm}
\end{eqnarray}
Here, $\mbox{\boldmath $g$}_{ij}$ is the Green's function 
determined by the unperturbed Hamiltonian ${\cal H}_1$ and 
density matrix $\widetilde{\rho}(-\infty)$ for 
the initial isolated system.
The Fourier transform  has been carried out 
for stationary states,
\begin{equation}
G(t_1,t_2) = \int_{-\infty}^{\infty} {d\omega \over 2\pi}
\  G(\omega) \,e^{-i\,\omega (t_1-t_2)} \;.
\label{eq:Fourier}
\end{equation}
For instance, in the noninteracting case ${\cal H}_C^U =0$,  
 the self-energy correction is caused only by 
 the couplings between the sample and reservoirs ${\cal H}_{\rm mix}$,
\begin{eqnarray}
&&
\!\!\!\!\!\!\!\!\!\!
\mbox{\boldmath $\Sigma$}_{ij}^{0} 
\,=\, - v_L 
         \left(\, 
                  \delta_{i,-1}\delta_{j,0} + \delta_{i,0}\delta_{j,-1}
          \,\right) 
\left[ 
          \matrix { 1 & \phantom{-}0  \cr   
                    0 & -1 \cr } 
                                \right] 
\nonumber \\
&& \quad \ \ 
 - v_R  \left(\, 
                  \delta_{l,N}\delta_{m,N+1} + \delta_{l,N+1}\delta_{m,N}
          \,\right) 
        \left[ 
          \matrix { 1 & \phantom{-} 0  \cr   
                    0 & -1 \cr } 
                                \right] .
\label{eq:self_mix}
\end{eqnarray}
Note that the four types Green's functions are not independent, 
\begin{eqnarray}
&&
\!\!\!\!\!\!\!\!\!\!\!\!\!\!\!\!\!
 G^{--} + \, G^{++} \, = \, 
 G^{+-} + \, G^{-+}  \;,
\qquad   
\Sigma^{--} + \Sigma^{++} \, = \,
 -  \, \Sigma^{-+} -  \,\Sigma^{+-} \;. 
\label{eq:depend}
\end{eqnarray}
Thus, the Dyson equation eq.\ (\ref{eq:Dyson}) can 
be expressed in terms of three independent quantities 
by carrying out a Unitary transformation 
$\mbox{\boldmath $P$}^{-1} \mbox{\boldmath $G$} \mbox{\boldmath $P$}$; 
\begin{eqnarray}
&&
\!\!\!\!\!\!\!\!\!\!\!\!\!\!\!\!\!
\mbox{\boldmath $P$}  =  {1 \over \sqrt{2} }
\left[ 
          \matrix { \phantom{-}1 & 1  \cr   
                              -1 & 1 \cr } 
                                \right] \;,
\\
&&
\!\!\!\!\!\!\!\!\!\!\!\!\!\!\!\!\!\!\!\!\!\!\!
\left[ 
 \matrix { 0 & G_{ij}^{a}   \cr
           G_{ij}^{r} & F_{ij}  \cr  }
                             \right]
 \,=\,  
\left[ 
 \matrix { 0 & g_{ij}^{a}   \cr
           g_{ij}^{r} & F^0_{ij}  \cr  }
                             \right] 
 \, +  
\sum_{lm}
\left[ 
 \matrix { 0 & g_{il}^{a}   \cr
           g_{il}^{r} & F^0_{il}  \cr  }
                             \right]  
\left[ 
 \matrix { \Omega_{lm} & \Sigma_{lm}^{r}   \cr
           \Sigma_{lm}^{a} & 0  \cr  }
                             \right] 
\left[ 
 \matrix { 0 & G_{mj}^{a}   \cr
           G_{mj}^{r} & F_{mj}  \cr  }
                             \right] .
\rule{0cm}{0.9cm}
\label{eq:Dyson2}
\end{eqnarray}
Here, $G^r$ and $G^a$ are the retarded and advanced Green's functions, 
respectively, 
\begin{eqnarray}
& & 
\!\!\!\!\!\!\!\!\!\!\!\!\!\!\!\!\!\!\!\!\!
G^{r} \,= \,G^{--} - \,G^{-+} \;, \quad \ \ 
G^{a} \, = \, G^{--} - \,G^{+-} \;, \quad \ \ 
F  \,=\, G^{--} + \,G^{++} , 
\label{eq:keldysh_to_gr}
\\
& &
\!\!\!\!\!\!\!\!\!\!\!\!\!\!\!\!\!\!\!\!\!
\Sigma^{r} \,=\, \Sigma^{--} + \,\Sigma^{-+} \;, \qquad 
\Sigma^{a} \,= \,\Sigma^{--} + \,\Sigma^{+-} \;, \qquad
\Omega   \,=\, \Sigma^{--} + \,\Sigma^{++} . 
\end{eqnarray}
The function $F$ and $\Omega$ link closely 
to a nonequilibrium distribution. 
Alternatively, 
the original four Green's functions can be expressed 
with these three functions, as
\begin{eqnarray}
G^{--} &=& [\,F + (G^r + G^a) \,]\,/2\;, \qquad  
G^{++} = [\,F - (G^r + G^a) \,]\,/2\;,
\label{eq:from_raf_to_keldysh1}
\\
G^{-+} &=& [\,F - (G^r - G^a) \,]\,/2\;, \qquad  
G^{+-} = [\,F + (G^r - G^a)\, ]\,/2\;. 
\label{eq:from_raf_to_keldysh2}
\end{eqnarray}
Similarly, the four self-energies are written 
in terms of   $\Sigma^r$, $\Sigma^a$ and $\Omega$,
\begin{eqnarray}
\Sigma^{--} &=& [\,\Omega + (\Sigma^r + \Sigma^a) \,]\,/2\;, \qquad  
\Sigma^{++} = [\,\Omega - (\Sigma^r + \Sigma^a) \,]\,/2\;,
\label{eq:from_raf_to_keldysh3}
\\
\Sigma^{-+} &=& -\,[\,\Omega - (\Sigma^r - \Sigma^a) \,]\,/2\;, \quad  
\Sigma^{+-} = -\,[\,\Omega + (\Sigma^r - \Sigma^a)\, ]\,/2\;. 
\label{eq:from_raf_to_keldysh4}
\end{eqnarray}

The Dyson equation for the three functions 
are deduced from eq.\ (\ref{eq:Dyson2}) 
\begin{eqnarray}
G^{r} &=& g^{r} + g^{r}\, \Sigma^{r} \,G^{r} \;, \qquad \quad 
G^{a} \ = \ g^{a} + g^{a}\, \Sigma^{a} \,G^{a} \;, 
\label{eq:Dyson3}
\\
F  &=& F^0 + F^0\, \Sigma^{a} \, G^{a} 
        + g^{r} \, \Sigma^{r} \, F + g^{r}\, \Omega \,G^{a}
\;.
\label{eq:Dyson4}
\end{eqnarray}
Here, we have suppressed the subscripts for simplicity,
and these equations should be understood symbolically.
Eq.\ (\ref{eq:Dyson4}) can be solved formally by 
using eq.\ (\ref{eq:Dyson3}), as
\begin{eqnarray}
F &=& \left[\,1- g^{r} \, \Sigma^{r} \, \right]^{-1} 
 F^0 \left[\, 1 + \Sigma^{a} \, G^{a}\,\right] \ + \ 
 \left[\,1- g^{r} \, \Sigma^{r} \, \right]^{-1} g^{r}\, \Omega \,G^{a} 
 \;, \nonumber
\\
   &=&  G^{r} \left\{ g^{r} \right\}^{-1} 
            F^0 \left\{ g^{a} \right\}^{-1} G^{a}
      \ + \ G^{r}\, \Omega \,G^{a} \;.
\label{eq:Dyson5}
\end{eqnarray}

\subsection{Green's function for the initial state} 
The unperturbed  Green's function
$\mbox{\boldmath $g$}_{ij}$ is determined 
by  ${\cal H}_1$ and the initial density matrix $\widetilde{\rho}(-\infty)$.
Initially at $t\to \infty$, 
the three regions are isolated and noninteracting.
Therefore, 
 $\mbox{\boldmath $g$}_{ij;\nu}$  for $\nu=L,\,R,\,C$ is given by 
\begin{eqnarray}
&&
\!\!\!\!\!\!\!\!\!\!\!\!\!\!\!\!\!
  g_{ij;\nu}^{r}(\omega) \,=\,  
    \sum_n { \phi_{n;\nu}(i) \phi^{*}_{n;\nu}(j)
              \over \omega - \epsilon_{n;\nu} + i \delta} \;,
\qquad 
  g_{ij;\nu}^{a}(\omega) \, = \,  
    \sum_n { \phi_{n;\nu}(i) \phi^{*}_{n;\nu}(j)
              \over \omega - \epsilon_{n;\nu} - i \delta} \;,
\label{eq:gra_0}
\\
&&
\!\!\!\!\!\!\!\!\!\!\!\!\!\!\!\!\!
F^{0}_{ij;\nu}(\omega) \,=\, 
[ 1 - 2\, f_{\nu}(\omega) ]  
  \left[\, g_{ij;\nu}^{r}(\omega) -  g_{ij;\nu}^{a}(\omega) \,\right]\;.
\rule{0cm}{0.7cm}
\label{eq:F_0}
\end{eqnarray}
Here,
$\epsilon_{n;\nu}$ and $\phi_{n;\nu}(i)$
are the one-particle eigenvalue and eigenstate of 
${\cal H}_{1;\nu}$. 
The information about the statistical distribution 
is contained in the function $F^{0}_{ij;\nu}$ via 
$f_{\nu}(\omega) = f(\omega - \mu_{\nu})$, where 
$f(\epsilon) = [ e^{\beta \epsilon} + 1 ]^{-1}$.
In the system we are considering, each of the reservoirs ($\nu=L,\, R$) 
has a continuous energy spectrum,  
and the isolated sample ($\nu=C$) has a discrete energy spectrum.
Thus, the full Green's function becomes to 
depend only on $\mu_{L}$ and $\mu_{R}$, 
and does not depend on $\mu_{C}$ \cite{HDW2}. 
This is because 
the contribution of $F^0$ to the corresponding full one 
$F$ arises in a sandwiched form 
 $\left\{ g^{r} \right\}^{-1} F^0 \left\{ g^{a} \right\}^{-1}$  
 as described in eq.\ (\ref{eq:Dyson5}). 
Thus, the singular contributions of $\delta$ functions in 
 $F^0 \propto  \left[\,g^{r} -  g^{a} \,\right]$ of the sample region  
  are canceled out by the zero points of the inverse Green's functions 
in both sides, to yield 
 $\left\{ g^{r} \right\}^{-1} F^0 \left\{ g^{a} \right\}^{-1} =0$ 
 for $\nu=C$. 



\subsection{Nonequilibrium current for noninteracting electrons} 
\label{subsec:current}

The nonequilibrium average of the charge and current 
can be deduced from the Green's functions.
For instance, 
by using eqs.\ (\ref{eq:G<}) and (\ref{eq:Fourier}),
an equal-time correlation function can be written in the form
\begin{equation}
\langle\, c^{\dagger}_{i \sigma} c^{\phantom{\dagger}}_{j \sigma} \,\rangle 
\, = \, -i G_{ji}^{-+}(0,0) \ = \  
  -i \int_{-\infty}^{\infty} {d\omega \over 2\pi}
 \  G_{ji}^{-+}(\omega) \;.  
\end{equation}
Therefore, the current flowing from the left lead to the sample 
is given by 
\begin{eqnarray}
&& 
J_L \,=\, i\,{\sl e} v_L \sum_{\sigma} \left[\,  
c^{\dagger}_{1 \sigma} c^{\phantom{\dagger}}_{0 \sigma} 
-c^{\dagger}_{0 \sigma} c^{\phantom{\dagger}}_{1 \sigma}\,\right] \;, 
\\
&&
\langle J_L \rangle 
\,=\,  2 {\sl e}  v_L 
\int_{-\infty}^{\infty} {d\omega \over 2\pi}
\, \left[\, G_{01}^{-+}(\omega) -  G_{10}^{-+}(\omega) \,\right]\;.  
\label{eq:current}
\end{eqnarray}
The expectation value for the current flowing from the sample to right lead,
$J_R$, can also be written in a similar form.
In the noninteracting case ${\cal H}_C^{U}=0$, 
eq.\ (\ref{eq:current}) can be 
rewritten in terms of retarded and advanced Green's functions
which link the two different interfaces of the sample \cite{Caroli};
\begin{eqnarray}
&&
\!\!\!\!\!\!\!\!\!\!\!\!
 \langle J\, \rangle \ 
 \,=\,  \ {2 {\sl e} \over h} \ \int_{-\infty}^{\infty} 
  d\omega \, 
\left[\, f_L(\omega) - f_R(\omega) \, \right]\, {\cal T}_0(\omega) \;,  
\label{eq:caro1}
\\
&&
\!\!\!\!\!\!\!\!\!\!\!\!
{\cal T}_0(\omega)  \,=\,  
4\  \Gamma_L(\omega) \, G_{1 N}^{a}(\omega) \, 
\Gamma_R(\omega) \, G_{N 1}^{r}(\omega) 
\;,
\rule{0cm}{0.7cm}
\label{eq:caro2}
\\
&&
\!\!\!\!\!\!\!\!\!\!\!\!
 \Gamma_L(\omega) \,=\, -\, \mbox{Im}  
                  \left[\, v_L^2 \, g_{0 0}^{r}(\omega) \,\right] \;,
\quad 
 \Gamma_R(\omega) 
\, = \, -\, \mbox{Im} \left[\,v_R^2 \, g_{N+1 N+1}^{r}(\omega)  \,\right]
\;.
\rule{0cm}{0.8cm}
\label{eq:caro3}
\end{eqnarray}
Note that 
$\langle J_L \rangle =\langle J_R \rangle$ 
($\equiv \langle J \,\rangle$) in steady states.
The outline of the derivation are provided 
in Sec.\ \ref{subsec:anderson_current} for a single Anderson impurity. 
Equation (\ref{eq:caro2}) implies that the current is determined by 
the electrons with the energy 
$\mu_R \lesssim \omega \lesssim \mu_L$ at low temperatures,
where $\mu_L-\mu_R =eV$.

For interacting electron systems,  
the nonequilibrium current can not generally 
be written in the form of eq.\ (\ref{eq:caro1}).
It does only in a particular case 
 where the connection of the two leads and 
 the sample has a special symmetry described by 
 a relation $\mbox{\boldmath $\Gamma$}^L(\epsilon) 
= \lambda\,\mbox{\boldmath $\Gamma$}^R(\epsilon)$ in 
the notation used in Ref.\ \cite{MW}.
In the interacting case,
the imaginary part of the self-energy caused 
by the inelastic scattering becomes finite. 
It links with the contributions of the vertex corrections,  
and the formulation becomes somewhat complicated.
Nevertheless, in the linear-response regime, 
the dc conductance for interacting electrons 
can be expressed in a Landauer-type form 
quite generally \cite{Landauer,Buttiker,FisherLee} 
even in the case without the special symmetry 
mentioned above \cite{ao10}. 
Specifically, at zero temperature $T=0$, the imaginary part 
of the self-energy and vertex corrections for the current 
become zero at the Fermi energy $\omega=0$, 
and the transmission probability can be written 
in the form of eq.\ (\ref{eq:caro2}) 
with the interacting Green's functions.
We discuss the details of these points in Sec.\ \ref{sec:Kubo}. 

\section{Out-of-equilibrium  Anderson model} 
\label{sec:Anderson_model}

\begin{figure}[tb]
\begin{center}

\setlength{\unitlength}{1mm}

\begin{picture}(100,20)
\thicklines

\put(15,4){\line(1,0){25}}
\put(15,16){\line(1,0){25}}
\put(40,4){\line(0,1){12}}

\put(66,4){\line(1,0){25}}
\put(66,16){\line(1,0){25}}
\put(66,4){\line(0,1){12}}

\multiput(41.0,10)(2,0){5}{\line(1,0){1}}
\multiput(56,10)(2,0){5}{\line(1,0){1}}

\put(53,10){\circle*{4}}

\put(43.0,12){\makebox(0,0)[bl]{\Large $v_L^{\phantom{\dagger}}$}}
\put(56.5,12){\makebox(0,0)[bl]{\Large $v_R^{\phantom{\dagger}}$}}
\put(50.5,0){\makebox(0,0)[bl]{\Large $\epsilon_d$}}

\end{picture}
\end{center}
\label{fig:Anderson_model}
\caption{Anderson impurity connected to two leads}
\end{figure}
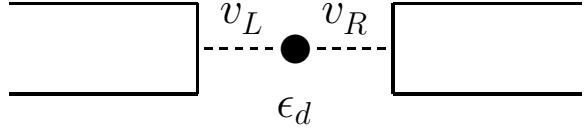

We now apply the Keldysh formalism 
to a single Anderson impurity connected to two leads  
as illustrated in Fig.\ \ref{fig:Anderson_model}.
It corresponds to the $N=1$ case of the Hamiltonian eq.\ (\ref{eq:Heq}),
and has been widely used as a model for quantum dots. 
For convenience,
we change the label for the sites: 
the new one for the impurity site is given by $0 \Rightarrow d$, 
and that for the interfaces at the left and right leads
are $0 \Rightarrow L$ and $N+1 \Rightarrow R$, respectively.

\subsection{Green's function for the Anderson impurity}

The self-energy in the interacting case 
can be classified into two parts. 
\begin{equation}
\mbox{\boldmath $\Sigma$}_{ij}(\omega) 
\,=\,
\mbox{\boldmath $\Sigma$}_{ij}^{0}(\omega) \, + \, 
\mbox{\boldmath $\Sigma$}_{U}(\omega)\, \delta_{i,d}\, \delta_{j,d}
\;.
\label{eq:Dyson_Anderson}
\end{equation}
Here,  $\mbox{\boldmath $\Sigma$}_{ij}^{0}$
corresponds the one defined in eq.\ (\ref{eq:self_mix}),
which represents the effects purely 
due to the mixing with reservoirs and sample. 
The remaining part $\mbox{\boldmath $\Sigma$}_{U}$ contains 
the contributions of the onsite Coulomb interaction $U$.
Substituting the self-energy eq.\ (\ref{eq:Dyson_Anderson})
into the Dyson equation (\ref{eq:Dyson}),  
we obtain a set of equations for the impurity Green's function,
\begin{eqnarray}
&&
    \mbox{\boldmath $G$}_{dd}(\omega) \,=\, 
    \mbox{\boldmath $g$}_{dd}(\omega)  
 + \mbox{\boldmath $g$}_{dd}(\omega)\,
\mbox{\boldmath $\Sigma$}_{U}(\omega)\,
\mbox{\boldmath $G$}_{dd}(\omega)         
\nonumber \\
& & \qquad \qquad \ \  -v_L \, \mbox{\boldmath $g$}_{dd}(\omega)\,
\mbox{\boldmath$\tau$}_3\,
\mbox{\boldmath $G$}_{Ld}(\omega) 
-v_R \, \mbox{\boldmath $g$}_{dd}(\omega)\,
\mbox{\boldmath$\tau$}_3\,
\mbox{\boldmath $G$}_{Rd}(\omega) 
\label{eq:Dyson_free_00}
         \;, 
           \rule{0cm}{0.5cm} 
          \\
  &&  \mbox{\boldmath $G$}_{Ld}(\omega) \,=\, 
-v_L \, 
\mbox{\boldmath $g$}_{L}(\omega)\,
\mbox{\boldmath$\tau$}_3\,
\mbox{\boldmath $G$}_{dd}(\omega) 
\label{eq:Dyson_free_-10}
         \;,         
        \rule{0cm}{0.6cm} 
        \\
   && \mbox{\boldmath $G$}_{Rd}(\omega) \,=\, 
-v_R \, \mbox{\boldmath $g$}_{R}(\omega)\,
\mbox{\boldmath$\tau$}_3\,
\mbox{\boldmath $G$}_{dd}(\omega) 
\label{eq:Dyson_free_10}
         \;.
        \rule{0cm}{0.6cm} 
\end{eqnarray}
where $g_{L}$ and $g_{R}$ are the Green's 
function at the interfaces at left and right, respectively.
The explicit form of the unperturbed Green's function 
at the impurity site is given by
 $\left\{ \mbox{\boldmath $g$}_{dd}(\omega) \right\}^{-1} 
 = 
 \left(\omega -\epsilon_d\right)
 \mbox{\boldmath$\tau$}_3$ with 
 one of the Pauli matrices   $\mbox{\boldmath$\tau$}_3$.
Substituting eqs.\ (\ref{eq:Dyson_free_-10}) and
(\ref{eq:Dyson_free_10}) into eq.\ (\ref{eq:Dyson_free_00}), 
we obtain
\begin{eqnarray}
    \mbox{\boldmath $G$}_{dd}(\omega) &=& 
    \mbox{\boldmath $g$}_{dd}(\omega)  
 + \mbox{\boldmath $g$}_{dd}(\omega)\,
\left[\, \mbox{\boldmath $\sigma$}(\omega)\,
 + 
\mbox{\boldmath $\Sigma$}_{U}(\omega)\,\right]\,
\mbox{\boldmath $G$}_{dd}(\omega)  \;,       
\\
\mbox{\boldmath $\sigma$}(\omega)
&=& 
v_L^2\, 
\mbox{\boldmath$\tau$}_3\,
\mbox{\boldmath $g$}_{L}(\omega)\, 
\mbox{\boldmath$\tau$}_3
\ + \ 
v_R^2\, 
\mbox{\boldmath$\tau$}_3\,
\mbox{\boldmath $g$}_{R}(\omega)\, 
\mbox{\boldmath$\tau$}_3
\;.
        \rule{0cm}{0.5cm} 
\label{eq:self_U=0}
\end{eqnarray}
Therefore,  
\begin{eqnarray}
    \left\{ \mbox{\boldmath $G$}_{dd}(\omega) \right\}^{-1} 
    &=& 
    \left\{ \mbox{\boldmath $G$}_{dd}^{(0)}(\omega) \right\}^{-1} 
        \ - \ 
\mbox{\boldmath $\Sigma$}_{U}(\omega)\;, 
\label{eq:Dyson_onsite}
\\
 \left\{ \mbox{\boldmath $G$}_{dd}^{(0)}(\omega) \right\}^{-1} 
    &=& 
    \left\{ \mbox{\boldmath $g$}_{dd}(\omega) \right\}^{-1} 
        \ - \ 
\mbox{\boldmath $\sigma$}(\omega)\;.
\label{eq:Dyson_onsite_free}
\end{eqnarray}
Here $\mbox{\boldmath $G$}_{dd}^{(0)}(\omega)$ 
is the Green's function for the noninteracting case. 
Furthermore, an alternatively form of 
the Dyson equation  $G  = g +  G \,\Sigma\, g$ yields
\begin{eqnarray}
    \mbox{\boldmath $G$}_{dL}(\omega) &=& 
-v_L \, 
\mbox{\boldmath $G$}_{dd}(\omega)\, 
\mbox{\boldmath$\tau$}_3\,
\mbox{\boldmath $g$}_{L}(\omega)\, 
\label{eq:Dyson_free_0-1}
\\
    \mbox{\boldmath $G$}_{dR}(\omega) &=& 
-v_R \, 
\mbox{\boldmath $G$}_{dd}(\omega) \,
\mbox{\boldmath$\tau$}_3\,
\mbox{\boldmath $g$}_{R}(\omega) \;.
\label{eq:Dyson_free_01}
\end{eqnarray}
Thus, the inter-site Green's functions can be deduced from  
$\mbox{\boldmath $G$}_{dd}(\omega)$ by using  
eqs.\ (\ref{eq:Dyson_free_-10})--(\ref{eq:Dyson_free_10})
and (\ref{eq:Dyson_free_0-1})--(\ref{eq:Dyson_free_01}).

The voltage-dependence arises via the unperturbed Green's functions 
for the leads at $\nu=L$ and $R$, 
\begin{eqnarray}
\mbox{\boldmath $g$}_{\nu}(\omega) &=& 
\mbox{\boldmath $P$} 
\left[ 
          \matrix { 0 & g_{\nu}^a(\omega)  \cr   
                    g_{\nu}^r(\omega) & F^0_{\nu}(\omega) \cr } 
                                \right] 
 \mbox{\boldmath $P$}^{-1}, 
\label{eq:gL_gR}
\\ 
F^0_{\nu}(\omega) &=& 
[ 1 - 2\, f_{\nu}(\omega) ]  
  [\, g_{\nu}^{r}(\omega) -  g_{\nu}^{a}(\omega) \,] 
  \;, \rule{0cm}{0.9cm}
   \nonumber
   \\
    &=&  -2i\, [1-2f_{\nu}(\omega)]\, \Gamma_{\nu}(\omega)/v_{\nu}^2
\label{eq:F0_11}
        \rule{0cm}{0.5cm} 
   \;,
\end{eqnarray}
where 
$\Gamma_{\nu}(\omega) 
 \equiv  -\,v_{\nu}^2\, \mbox{Im} \left[ g_{\nu}^{r}(\omega) \right]$.
The pure mixing part of 
the self energy $\mbox{\boldmath $\sigma$}(\omega)$ can be 
calculated from  eqs.\ (\ref{eq:self_U=0}) and (\ref{eq:gL_gR}), 
\begin{eqnarray}
 \mbox{\boldmath $\sigma$}(\omega)
 &=& 
 \mbox{\boldmath $P$} \, 
\left[ 
 \matrix { \Omega^{(0)}(\omega) & \sigma^{r}(\omega)   \cr
           \sigma^{a}(\omega) & 0  \cr  }
                             \right] 
 \mbox{\boldmath $P$}^{-1}
\;,
\\
\Omega^{(0)}(\omega) &=& 
    v_L^2\,F^0_{L}(\omega) + v_R^2\,F^0_{R}(\omega) 
    \;,
\label{eq:self_U=0_omega}
\rule{0cm}{0.6cm}
\\
\sigma^{r}(\omega) &=& v_L^2\, g_{L}^r(\omega) 
+ v_R^2\, g_{R}^r(\omega) \;,
        \rule{0cm}{0.5cm} 
\label{eq:self_U=0_ra} 
\end{eqnarray}
and $\sigma^{a}(\omega)=\left\{ \sigma^{r}(\omega) \right\}^*$.
Then the noninteracting Green's 
function $\mbox{\boldmath $G$}_{dd}^{(0)}(\omega)$ 
 can be determined via eq.\ (\ref{eq:Dyson_onsite_free}),  
\begin{eqnarray}
G_{dd}^{(0)--}(\omega) &=&  
\left[\,1-\,f_{\rm eff}(\omega)\,\right] G_{dd}^{(0)r}(\omega) \,
+\,f_{\rm eff}(\omega)
        \,G_{dd}^{(0)a}(\omega) \;, 
\label{eq:G0_00^--}
\\ 
G_{dd}^{(0)++}(\omega) &=&  
-\,f_{\rm eff}(\omega)
\,G_{dd}^{(0)r}(\omega) \,
-\,\left[\,1-\,f_{\rm eff}(\omega)\,\right]
        G_{dd}^{(0)a}(\omega) \;, 
 \label{eq:G0_00^++}
\\ 
G_{dd}^{(0)-+}(\omega) &=&  
-\,f_{\rm eff}(\omega)
        \left[\,G_{dd}^{(0)r}(\omega)- G_{dd}^{(0)a}(\omega)\,\right] , 
\label{eq:G0_00^-+}
\\ 
G_{dd}^{(0)+-}(\omega) &=&  
\left[\,1-\,f_{\rm eff}(\omega)\,\right]
        \left[\,G_{dd}^{(0)r}(\omega)- G_{dd}^{(0)a}(\omega)\,\right] ,
\label{eq:G0_00^+-}  
\end{eqnarray}
where
\begin{eqnarray}
&&f_{\rm eff}(\omega) \,=\,
{  f_L(\omega) \,\Gamma_L 
  + f_R(\omega) \,\Gamma_R
  \over 
 \Gamma_L +\Gamma_R } \;,
\label{eq:f_0}
\\
&& G_{dd}^{(0)r}(\omega) \,=\,
\frac{1}{\omega-\epsilon_d -\sigma^r(\omega)} \;,
\rule{0cm}{0.6cm}
\end{eqnarray}
and $G_{dd}^{(0)a}(\omega)=\left\{ G_{dd}^{(0)r}(\omega) \right\}^*$. 
Thus, the effects of the bias voltage arise through the distribution function 
$f_{\rm eff}(\omega)$.

The full Green's function  for the impurity site 
can be expressed, using eqs.\ (\ref{eq:Dyson3})--(\ref{eq:Dyson5})
and (\ref{eq:F0_11})--(\ref{eq:self_U=0_omega}), as
\begin{eqnarray}
G_{dd}^{r}(\omega) &=& {1 \over \omega - \epsilon_d 
 -\sigma^r(\omega) -\Sigma_U^r(\omega)}\;,
\label{eq:Gr_00}
\\
F_{dd}(\omega) &=&
G_{dd}^{r}(\omega)\, \left[\,
\Omega^{(0)}(\omega)\,+ \, 
\Omega_U(\omega)\,\right]\, 
G_{dd}^{a}(\omega)\;, 
\rule{0cm}{0.6cm}
\label{eq:F_00}
\end{eqnarray}
where $G_{dd}^{a}(\omega)=\left\{ G_{dd}^{r}(\omega) \right\}^*$. 
Note that 
$\Omega_U(\omega) = -\Sigma_U^{-+}(\omega)-\Sigma_U^{+-}(\omega)$ 
and $F_{dd}(\omega)$ are  pure imaginary.  
The four elements of $\mbox{\boldmath $G$}_{dd}(\omega)$ can be written,  
using 
eqs.\ (\ref{eq:from_raf_to_keldysh1})--(\ref{eq:from_raf_to_keldysh2})   
and (\ref{eq:Gr_00})--(\ref{eq:F_00}),  
as
\begin{eqnarray}
G_{dd}^{--}(\omega) &=&  
\left[\,1-\,\widetilde{f}_{\rm eff}(\omega)\,\right] G_{dd}^{r}(\omega) \,
+\,\widetilde{f}_{\rm eff}(\omega)
        \,G_{dd}^{a}(\omega) \;, 
\label{eq:G_00^--}
\\ 
G_{dd}^{++}(\omega) &=&  
-\,\widetilde{f}_{\rm eff}(\omega)
\,G_{dd}^{r}(\omega) \,
-\,\left[\,1-\,\widetilde{f}_{\rm eff}(\omega)\,\right]
        G_{dd}^{a}(\omega) \;, 
 \label{eq:G_00^++}
\\ 
G_{dd}^{-+}(\omega) &=&  
-\,\widetilde{f}_{\rm eff}(\omega)
        \left[\,G_{dd}^{r}(\omega)- G_{dd}^{a}(\omega)\,\right] , 
\label{eq:G_00^-+}
\\ 
G_{dd}^{+-}(\omega) &=&  
\left[\,1-\,\widetilde{f}_{\rm eff}(\omega)\,\right]
        \left[\,G_{dd}^{r}(\omega)- G_{dd}^{a}(\omega)\,\right] , 
\label{eq:G_00^+-} 
\end{eqnarray}
where $G_{dd}^{++}(\omega) = - \{ G_{dd}^{--}(\omega) \}^*$,
and $\widetilde{f}_{\rm eff}(\omega)$ is a correlated 
distribution defined by 
\begin{eqnarray}
\widetilde{f}_{\rm eff}(\omega) &=&
{ f_L(\omega) \,\Gamma_L 
  + f_R(\omega) \,\Gamma_R - 
  {\displaystyle \mathstrut 1\over \displaystyle \mathstrut 2i}\,
  \Sigma_{U}^{-+}(\omega)
  \over 
 \Gamma_L +\Gamma_R - {\rm Im} \Sigma_{U}^r(\omega)} \;.
\label{eq:f_int}
\end{eqnarray}
With this distribution function, 
the number of the electrons in the impurity site can be written in the form
\begin{eqnarray}
\langle n_d \rangle &=& 
2\,\int  
  d\omega\, \widetilde{f}_{\rm eff}(\omega) 
  \left(-\,{1 \over\pi}\right)  {\rm Im}\,G_{dd}^{r}(\omega)  
\;.
\end{eqnarray}
In the equilibrium case $\mu \equiv \mu_L = \mu_R$,
both $\widetilde{f}_{\rm eff}(\omega)$ and $f_{\rm eff}(\omega)$
coincide with the Fermi function $f(\omega)$, 
because of the property eq.\ (\ref{eq:self_lesser_U}).

\subsection{Properties of the Green's functions at $eV=0$} 

We summarize here the properties of the Keldysh Green's function 
in the limit of the zero-bias voltage $eV=0$, at which $\mu_L=\mu_R$. 
In this case, 
the four self-energies can be written in the form 
\begin{eqnarray}
\Sigma_{U:{\rm eq}}^{--}(\omega) &=& 
\left[\,1- f(\omega)\, \,\right] \,
\Sigma_{U:{\rm eq}}^r(\omega) 
\, + \, f(\omega)\,\Sigma_{U:{\rm eq}}^a(\omega) 
\;, 
\\
\Sigma_{U:{\rm eq}}^{++}(\omega) &=& 
-f(\omega)\,\Sigma_{U:{\rm eq}}^r(\omega) 
\, - \, 
\left[\,1- f(\omega)\, \,\right] \,
\Sigma_{U:{\rm eq}}^a(\omega) 
\;, 
\\
\Sigma_{U:{\rm eq}}^{-+}(\omega) &=& 
f(\omega)\, \left[\, 
\Sigma_{U:{\rm eq}}^r(\omega) - \Sigma_{U:{\rm eq}}^a(\omega) 
\,\right] \;, 
\label{eq:self_lesser_U}
\\
\Sigma_{U:{\rm eq}}^{+-}(\omega) &=&
-\, \left[\,1-f(\omega)\,\right]\, \left[\, 
\Sigma_{U:{\rm eq}}^r(\omega) - \Sigma_{U:{\rm eq}}^a(\omega) 
\,\right] \;. 
\label{eq:self_bigger_U}
\end{eqnarray}
Furthermore, in equilibrium,
$F_{dd:{\rm eq}}(\omega)$ and
$\Omega_{U:{\rm eq}}(\omega)$ are determined by
the retarded and advanced functions,
\begin{eqnarray}
F_{dd:{\rm eq}}(\omega) &=& 
 \left[\,1 - 2 f(\omega)\,\right]\, \left[\, 
G_{dd:{\rm eq}}^r(\omega) - G_{dd:{\rm eq}}^a(\omega) 
\,\right] \;,
\\
\Omega_{U:{\rm eq}}(\omega) &=& 
 \left[\,1 - 2 f(\omega)\,\right]\, \left[\, 
\Sigma_{U:{\rm eq}}^r(\omega) - \Sigma_{U:{\rm eq}}^a(\omega) 
\,\right] \;.
\end{eqnarray}

Specificality at zero temperature,  
 $\Sigma_{U:{\rm eq}}^{-+}(\omega)$ 
 and $\Sigma_{U:{\rm eq}}^{+-}(\omega)$ vanish, 
respectively, at $\omega>\mu$ and $\omega<\mu$,
because of the Fermi function  
in eqs.\ (\ref{eq:self_lesser_U}) and (\ref{eq:self_bigger_U}).
Similarly, the Green's functions 
$G_{dd:{\rm eq}}^{-+}(\omega)$ and $G_{dd:{\rm eq}}^{+-}(\omega)$ 
also vanish at $\omega>\mu$ and $\omega<\mu$, respectively.
Therefore, at the equilibrium ground state, 
the usual $T=0$ formalism which yields 
a single-component Dyson equation 
\begin{eqnarray}
G_{dd:{\rm eq}}^{--}(\omega) &=& 
G_{dd:{\rm eq}}^{(0)--}(\omega) +
G_{dd:{\rm eq}}^{(0)--}(\omega)\,
\Sigma_{U:{\rm eq}}^{--}(\omega)\,
G_{dd:{\rm eq}}^{--}(\omega) 
\end{eqnarray}
becomes available.

\subsection{Current through the Anderson impurity} 
\label{subsec:anderson_current}

The operators for the 
current flows from the left reservoir to the sample $J_L$ 
and that from the sample to the right reservoir $J_R$ 
are given by
\begin{eqnarray}
&& J_L \, =\, i\,{\sl e} \sum_{\sigma}\, v_L \left[\,  
d^{\dagger}_{\sigma} c^{\phantom{\dagger}}_{L \sigma} 
-c^{\dagger}_{L \sigma} d^{\phantom{\dagger}}_{\sigma}\,\right] ,
\label{eq:current_L}
\\
&& J_R \,=\, i\,{\sl e}  \sum_{\sigma}\, v_R \left[\,  
c^{\dagger}_{R \sigma} d^{\phantom{\dagger}}_{\sigma} 
-d^{\dagger}_{\sigma} c^{\phantom{\dagger}}_{R \sigma}\,\right] .
\end{eqnarray}
As mentioned in Sec.\ \ref{subsec:current},
the expectation values  can be expressed in the form 
\begin{eqnarray}
&& \langle J_L \rangle 
\,=\,  2 {\sl e}  
\int_{-\infty}^{\infty} {d\omega \over 2\pi}
\ v_L \left[\, G_{Ld}^{-+}(\omega) -  G_{dL}^{-+}(\omega) \,\right]\;,
 \rule{0cm}{0.6cm}
\label{eq:current_L}
\\
&&\langle J_R \rangle 
\,=\,  2 {\sl e}   
\int_{-\infty}^{\infty} {d\omega \over 2\pi}
\ v_R \left[\, G_{dR}^{-+}(\omega) -  G_{Rd}^{-+}(\omega) \,\right]\;.
\label{eq:current_R}
\end{eqnarray}
  
The inter-site Green's functions in the right-hand side 
of eq.\ (\ref{eq:current_R}) can be expressed  
in terms of $\mbox{\boldmath $G$}_{dd}(\omega)$  
using 
 eqs.\ (\ref{eq:Dyson_free_-10})--(\ref{eq:Dyson_free_10}),
and 
 (\ref{eq:Dyson_free_0-1})--(\ref{eq:Dyson_free_01}), as
\begin{eqnarray}
 &&
\!\!\!\!\! \!\!\!\!\!\!\!\!\!\!\!\!
 \mbox{\boldmath $P$}^{-1}\,
 \mbox{\boldmath $G$}_{Rd}
  \mbox{\boldmath $P$} 
 \,=\, 
\left[ 
 \matrix { 0 & G_{Rd}^{a}  \cr
           G_{Rd}^{r} & F_{Rd}  \cr  }
                             \right] 
\, = \, -v_R
\left[ 
 \matrix { 0 & g_{R}^{a}\, G_{dd}^{a}   \cr
           g_{R}^{r}\, G_{dd}^{r} \quad  &  
           g_{R}^{r}\,F_{dd} + F_{R}^{0}\,G_{dd}^{a}  \cr  }
                             \right] , 
\label{eq:G_10}
\\ 
 && 
\!\!\!\!\! \!\!\!\!\!\!\!\!\!\!\!\!
 \mbox{\boldmath $P$}^{-1}\,
 \mbox{\boldmath $G$}_{dR}
  \mbox{\boldmath $P$} 
 \,=\,
\left[ 
 \matrix { 0 & G_{dR}^{a}  \cr
           G_{dR}^{r} & F_{dR}  \cr  }
                             \right] 
\, = \, -v_R
\left[ 
 \matrix { 0 & G_{dd}^{a}\, g_{R}^{a}   \cr
           G_{dd}^{r}\, g_{R}^{r} \quad  &  
           G_{dd}^{r}\,F_{R}^{0} + F_{dd}\,g_{R}^{a}  \cr  }
                             \right] . 
\label{eq:G_01}
\end{eqnarray}
Thus, using 
eqs.\ (\ref{eq:from_raf_to_keldysh2}), (\ref{eq:F_00}), 
 and (\ref{eq:G_10})--(\ref{eq:G_01}), 
we obtain 
\begin{eqnarray}
&&
\!\!\!\!\!\!\!\!\!\!\!\!\!\!\!
v_R\, [G_{dR}^{-+} - G_{Rd}^{-+}] \,=\, 
{v_R \over 2}\,[F_{dR} - F_{Rd}]  
\nonumber \\
 && \ \ \ \qquad \qquad =\,  
 {v_R^2 \over 2}
 \left[\,(g_{R}^r-g_{R}^a)\,F_{dd} 
 - F^0_{R}\,(G_{dd}^r-G_{dd}^a) \,\right] 
 \nonumber \\
&& \ \ \ \qquad \qquad = \,
\Gamma_R\, G_{dd}^r\, G_{dd}^a\, 
\bigl[\,
      4\Gamma_L\, (f_L - f_R) 
      -i\, \Omega_U
-2 (1-2f_R)\, {\rm Im} \Sigma^r_U \,\bigr] .
\nonumber \\ 
\label{eq:JR_ker}
\end{eqnarray}
Similarly, for the right-hand side of eq.\ (\ref{eq:current_L}),
we obtain 
\begin{eqnarray}
&&
\!\!\!\!\!\!\!\!\!\!\!\!\!\!\!
v_L\, [G_{Ld}^{-+} - G_{dL}^{-+}] \,=\, 
{v_L \over 2}\,[F_{Ld} - F_{dL}] 
\nonumber \\
 && \ \ \ \qquad \qquad =\,  
 {v_L^2 \over 2}
 \left[\,-(g_{L}^r-g_{L}^a)\,F_{dd} 
 + F^0_{L}\,(G_{dd}^r-G_{dd}^a) \,\right] 
 \nonumber \\
&& \ \ \ \qquad \qquad = \,
\Gamma_L\, G_{dd}^r\, G_{dd}^a\, 
\bigl[\,
      4\Gamma_R\, (f_L - f_R) 
      +i\, \Omega_U
+2 (1-2f_L)\, {\rm Im}\, \Sigma^r_U \,\bigr] . 
\nonumber \\ 
\label{eq:JL_ker}
\end{eqnarray}
If the two couplings between the Anderson impurity and leads 
have a property $\Gamma_L(\omega) = \lambda \Gamma_R(\omega)$
with $\lambda$ being a constant \cite{MW},
the expression for current can be simplified 
by taking an average,
$\Gamma_L \times$(\ref{eq:JR_ker}) + $\Gamma_R \times$(\ref{eq:JL_ker}), as
\begin{eqnarray}
&&
\!\!\!\!\!\!\!\!\!\!\!
\langle J\, \rangle \, = \, 
\frac{ 
\Gamma_L\langle\, J_R\, \rangle 
+ \Gamma_R\langle\, J_L\, \rangle}{\Gamma_R + \Gamma_L} 
\nonumber \\
 && 
\!\!\!\!\!\!\!\!\!\!\!
 \qquad
 \, = \,
   {2 {\sl e} \over h} \ \int_{-\infty}^{\infty} 
  d\omega\, 
   \left[\, f_L(\omega) - f_R(\omega) \, \right]\, 
\frac{4\, \Gamma_L \Gamma_R}{\Gamma_R + \Gamma_L} 
\,\left[\, - {\rm Im}\, G_{dd}^r(\omega) \,\right]
\;.
\rule{0cm}{0.7cm}
\label{eq:caroli}
\end{eqnarray}
Note that 
$\langle J \rangle=\langle J_L \rangle =\langle J_R \rangle$ 
for stationary states.

\subsection{Perturbation expansion with respect to ${\cal H}^U_C$} 

So far, we have discussed general properties of  
the nonequilibrium Green's functions. 
To study how the inter-electron interactions affect 
the transport properties, 
the self-energy $\mbox{\boldmath $\Sigma$}_{U}(\omega)$ 
must be calculated with reliable methods. 
Because the noninteracting Green's function, 
which includes the couplings between the leads and the Anderson impurity,
have already been obtained in eqs.\ (\ref{eq:G0_00^--})--(\ref{eq:G0_00^-+}), 
the remaining task is 
calculating $\mbox{\boldmath $\Sigma$}_{U}(\omega)$, for instance, 
by taking $\mbox{\boldmath $G$}_{dd}^{(0)}$ to be 
the unperturbed Green's function. 
Then, the interacting Green's function $\mbox{\boldmath $G$}_{dd}$ 
are deduced via the Dyson equation (\ref{eq:Dyson_onsite}).

The self-energy $\mbox{\boldmath $\Sigma$}_{U}(\omega)$ 
can be calculated with the perturbation expansion 
with respect to the inter-electron interaction ${\cal H}_C^U$.
For generating the perturbation series,  
it is convenient to introduce an effective action,
 \begin{eqnarray}
    S(\eta^{\dagger}, \eta) &=&  
   S_0(\eta^{\dagger}, \eta) 
   \ + \ 
   S_U(\eta^{\dagger}, \eta)
    \ + \  
   S_{\rm ex}(\eta^{\dagger}, \eta) \;,
    \label{eq:action_tot}
   \\
   S_0(\eta^{\dagger}, \eta) &=&
   \sum_{\sigma} \int \! dt\,dt'\ 
 \mbox{\boldmath $\eta$}_{\sigma}^{\dagger}(t)\, 
       \mbox{\boldmath $K$}^{(0)}_{dd}(t,t')\, 
 \mbox{\boldmath $\eta$}_{\sigma}(t')\, 
 \label{eq:action_0} \;,
 \rule{0cm}{0.7cm}
 \\
   S_U(\eta^{\dagger}, \eta) &=&
   -U \int \!\! dt \,
 \Bigl[ \, 
 \eta_{\uparrow -}^{\dagger}(t)
 \eta_{\uparrow -}(t)
 \eta_{\downarrow -}^{\dagger}(t)
 \eta_{\downarrow -}(t) 
 \nonumber \\
& & \qquad \qquad
 \, - \,   
 \eta_{\uparrow +}^{\dagger}(t)
 \eta_{\uparrow +}(t)
 \eta_{\downarrow +}^{\dagger}(t)
 \eta_{\downarrow +}(t) \,\Bigr] \;. 
 \label{eq:action_U}
 \end{eqnarray}
 Here, 
 $
 \mbox{\boldmath $\eta$}_{\sigma}^{\dagger}(t)
 = \left(\, 
 \eta_{\sigma -}^{\dagger}(t), \, 
 \eta_{\sigma +}^{\dagger}(t) \,\right) 
 $ is a two-component Grassmann number corresponding to the $-$ and $+$ 
 branches of the Keldysh contour shown in Fig.\ \ref{fig:loop}.
 The Kernel $\mbox{\boldmath $K$}^{(0)}_{dd}$ is determined by  
 the noninteracting Greens function,
 \begin{eqnarray}
   \mbox{\boldmath $K$}^{(0)}_{dd}(\omega)  &\equiv&
    \left\{\mbox{\boldmath $G$}_{dd}^{(0)}(\omega)\right\}^{-1}
 \;, \\
    \mbox{\boldmath $K$}^{(0)}_{dd}(t,t') &=&
    \int {d\omega \over 2 \pi}\,
    \mbox{\boldmath $K$}^{(0)}_{dd}(\omega) 
    \, e^{-i\omega (t-t')} 
        . 
 \end{eqnarray}
 In eq.\ (\ref{eq:action_U}), 
 the sign for the interaction along the $-$ branch 
 and that for $+$ branch are different. 
 This correspond to the sign arises in eq.\ (\ref{eq:O_H_expand}),
 and it is determined from which of
 the time-evolution operators, $U(+\infty,-\infty)$ or $U(-\infty,+\infty)$, 
 the perturbation terms arise.
For $S_{\rm ex}(\eta^{\dagger},\,\eta)$ in eq.\ (\ref{eq:action_tot}),
we introduce an external source of two anticomutating c-numbers 
$
 \mbox{\boldmath $j$}_{\sigma}^{\dagger}(t)
 = \left(\, 
 j_{\sigma -}^{\dagger}(t), \, 
 j_{\sigma +}^{\dagger}(t) \,\right) 
 $ following along the standard procedure \cite{PathIntegral}, 
\begin{eqnarray}
   S_{\rm ex}(\eta^{\dagger},\,\eta) &=& 
   -\sum_{\sigma} \int \! dt\,
   \left[\,  
 \mbox{\boldmath $\eta$}_{\sigma}^{\dagger}(t)\, 
       \mbox{\boldmath $j$}_{\sigma}(t) \ + \  
       \mbox{\boldmath $j$}_{\sigma}^{\dagger}(t)\, 
 \mbox{\boldmath $\eta$}_{\sigma}(t')
\,\right]
 \label{eq:action_ex} \;.
\end{eqnarray}
In this formulation, the Green's functions are generated from  
a functional $Z[j]$, as
\begin{eqnarray}
Z[j] 
&\equiv& 
\int 
D\eta^{\dagger} D\eta \  e^{iS(\eta^\dagger,\,\eta)}
\;, 
\label{eq:Green_path}
\\
G^{\nu\nu'}_{dd,\sigma}(t,t') &=&
-i\,{1 \over Z[0]}\, 
\left.
{\delta \over \delta j^{\dagger}_{\sigma\nu}(t)}\ 
{\delta \over \delta j^{\phantom{\dagger}}_{\sigma\nu'}(t')}\ 
Z[j] \,\right|_{j=0} 
\;.
\end{eqnarray}
In the noninteracting case,
the functional integration can be calculated analytically
\begin{eqnarray}
Z^{(0)}[j] 
&\equiv& 
\int 
D\eta^{\dagger} D\eta \  
e^{i\,\left[ S_0(\eta^\dagger,\,\eta) 
\,+\, S_{\rm ex}(\eta^\dagger,\,\eta) \right]}
\;, 
\nonumber\\
&=& Z^{(0)}[0]\  \exp\left[ -i
 \sum_{\sigma} \int \! dt\,dt'\,
 \mbox{\boldmath $j$}_{\sigma}^{\dagger}(t)\, 
       \mbox{\boldmath $G$}^{(0)}_{dd}(t,t')\, 
 \mbox{\boldmath $j$}_{\sigma}(t')\, 
\right] .
\label{eq:Z0}
\end{eqnarray}
The generating functional $Z[j]$ can  be rewritten 
in the form  
\begin{eqnarray}
Z[j] &=&  e^{
  i S_U\left(-i {\delta \over\delta j},\, i {\delta \over \delta j^{\dagger}}
  \right) } \, Z^{(0)}[j]  \;.
  \label{eq:Z_vs_Z0}
\end{eqnarray}
Here, $\eta$ and $\eta^{\dagger}$ in the action
$S_U(\eta^{\dagger},\eta)$ has been replaced 
by the functional derivatives 
\begin{equation} 
\eta^{\dagger}_{\sigma\nu}(t) \, \Rightarrow 
-i\, {\delta \over\delta j_{\sigma\nu}(t)}\;, 
\qquad 
\eta^{\phantom{\dagger}}_{\sigma\nu}(t) \, \Rightarrow 
i\, {\delta \over\delta j_{\sigma\nu}^{\dagger}(t)}\;.
\end{equation} 
The perturbation series  can be  obtained by substituting 
eq.\ (\ref{eq:Z0}) into (\ref{eq:Z_vs_Z0}) 
and then expanding $e^{iS_U}$ in a power series of $S_U$.

\subsection{Fermi-liquid behavior at low bias voltages} 

The out-of-equilibrium Anderson model has been 
studied by a number of theoretical approaches.
In this section, we discuss briefly 
the low-bias behavior of the Green's function 
and differential conductance,  
which have been deduced from the Ward identities 
for the Keldysh formalism \cite{ao_neq}.
In equilibrium and linear-response regime,
the low-energy properties at $\max(\omega, T) \ll T_K$
can be described by the local Fermi liquid theory,
where $T_K$ is the Kondo temperature \cite{Hewson}.    
The results deduced from the Ward identities show that
the nonlinear properties at small bias-voltages $eV \ll T_K$ 
can also be described by the local Fermi liquid theory.

The low-energy behavior of $\mbox{Im}\, \Sigma^r_U(\omega)$ 
has been calculated exactly 
up to terms of order $\omega^2$, $(eV)^2$, and $T^2$,    
\begin{eqnarray}
&&
\!\!\!\!\!\!\!\!\!\!\!\!\!\!\!\!\!\!\!\!
\mbox{Im}\, \Sigma^r_U(\omega) 
  \ = \    
   \, -\, 
  { \pi \over 2 } \left\{A_{\rm eq}(0)\right\}^3\, \left| 
  \Gamma_{\uparrow\downarrow;\downarrow\uparrow}(0,0;0,0)\right|^2
\,
\nonumber \\
 && \qquad \qquad  \times 
        \left[\,
            \left(\,\omega - 
              \alpha\, {\sl e}V\, 
              \right)^2 
              +  { 3\,\Gamma_L \Gamma_R 
                 \over \left( \Gamma_L + \Gamma_R \right)^2}
                \,(eV)^2 
              +(\pi T)^2  
           \,\right] , 
\label{eq:self_imaginary}
\end{eqnarray}
where $\Gamma_{\sigma\sigma';\sigma'\sigma}(\omega,\omega';\omega',\omega)$ 
is the vertex function for the causal Green's function
in the zero-temperature formalism, and
 $A_{{\rm eq}}(\omega) 
= -{\rm Im}\, G^{r}_{dd:{\rm eq}}(\omega)/\pi$. 
The parameter $\alpha$ is defined by 
$ \alpha \equiv (\alpha_L \Gamma_L - \alpha_R \Gamma_R)/ 
 (\Gamma_L+ \Gamma_R) $, where
$\alpha_L$ and $\alpha_R$ are constants which have been introduced 
to specify how the bias voltage is applied to the equilibrium state.
Namely,  $\mu_L \equiv \alpha_L \,{\sl e}V$ and
$\mu_R \equiv - \alpha_R \,{\sl e}V$ with $\alpha_L + \alpha_R=1$.

The real part of the self-energy is generally complicated.
However, it is simplified in the electron-hole symmetric case 
for   
$\epsilon_d = -U/2$, $\Gamma_L = \Gamma_R$, and $\mu_L=-\mu_R=eV/2$. 
 In this case the spectral wight at the Fermi energy becomes 
 $A_{\rm eq}(0) = 1/(\pi\Delta)$  with $\Delta=\Gamma_L+\Gamma_R$,
 and the low-energy behavior of the 
 real part of the self-energy is given by     
\begin{eqnarray}
&& \mbox{Re}\, \Sigma^r_U(\omega) \, = \, \left( 1 - z^{-1} \right)
 \omega \, + \, O(\omega^3) \;,
\label{eq:re_S}
\\
&& z^{-1} \ \equiv \  
1- \left.{\partial \Sigma_{U:{\rm eq}}^r(\omega) \over \partial \omega }
\right|_{\omega=0}  \;.
\rule{0cm}{1cm}
\label{eq:wave_remorm}
\end{eqnarray}
Here, the constant Hartree term $U/2$ is included into 
the unperturbed part, and it set the position of 
the Kondo peak on the Fermi energy $\epsilon_d +U/2=0$.
Therefore,  $G^r(\omega)$ can be deduced 
exactly up to terms of order $\omega^2$, $T^2$ and $(eV)^2$ 
from eqs.\ (\ref{eq:self_imaginary}) and (\ref{eq:re_S}), 
\begin{equation}
G^r(\omega) \ \simeq \ 
{z \over 
\omega + i \, \widetilde{\Delta} 
      +\,i \, 
 {\displaystyle \mathstrut \widetilde{U}^2 
 \over \displaystyle 
 2 \widetilde{\Delta} 
 \rule{0cm}{0.4cm} \mathstrut 
       (\pi \widetilde{\Delta})^2}
\left[ \, \omega^2 + 
{\displaystyle \mathstrut 3  \over \displaystyle \mathstrut 4}
\,({\sl e}V)^2 +(\pi T)^2
       \right]
       }         ,
 \label{eq:Gr_symmetric_case}
\end{equation}
where the renormalized parameters are defined by 
\begin{equation}
\widetilde{\Delta}\, \equiv \,z \Delta \;, \qquad
\widetilde{U} \, \equiv\,  z^2\, 
\Gamma_{\uparrow\downarrow;\downarrow\uparrow}(0,0;0,0) \;.
\label{eq:renorm_parm}
\end{equation}
The  order $U^2$ result \cite{HDW2} 
can be reproduced from eq.\ (\ref{eq:Gr_symmetric_case})
by replacing $\widetilde{U}$ with the bare Coulomb interaction $U$ 
and using the perturbation result for the renormalization factor 
$z = 1 - (3-\pi^2/4) \,u^2 + \cdots$, where $u= U/(\pi \Delta)$.

This result shows that in the symmetric case the low-voltage behavior 
is characterized by the two parameters $\widetilde{\Delta}$  
and $\widetilde{U}$. 
These two parameters  
are defined with respect to the equilibrium ground state,
and the exact Bethe ansatz results exist for these parameters.  
The width of the Kondo resonance $\widetilde{\Delta}$ decreases 
with increasing $U$, 
and becomes close to $\widetilde{\Delta} \simeq (4/\pi) T_K$     
for large $U$ with the Kondo temperature defined by 
\begin{equation}
T_K \, = \, \pi \Delta \sqrt{u / (2 \pi)}\, \exp[-\pi^2 u/8 + 1/(2u)] \;.
\label{eq:T_K}
\end{equation}
The Wilson ratio is usually defined by 
$R \equiv \widetilde{\chi}_s/\widetilde{\gamma}$,
where $\widetilde{\gamma}$ and $\widetilde{\chi}_s$ are  
the enhancement factors for the $T$-linear specific heat 
and spin susceptibility,  respectively \cite{YY}.
Alternatively, it corresponds to the ration of
$\widetilde{U}$ to $\widetilde{\Delta}$,
\begin{equation}
R -1 \, = \,  \widetilde{U}/(\pi \widetilde{\Delta}) \;.
\label{eq:wilson_vs_renorm}
\end{equation}
The Wilson ratio takes a value $R=1$ for $U=0$, 
and it reaches  $R = 2$ in the strong-coupling limit $U \to \infty$.

The nonequilibrium current $\langle J \rangle$ is calculated 
by substituting eq.\ (\ref{eq:Gr_symmetric_case}) into eq.\ (\ref{eq:caroli}).
Then, the differential conductance $dJ/dV$ are determined 
exactly up to terms of order $T^2$ and $(eV)^2$,  
\begin{equation}
 {dJ \over dV}  
\ = \ {2 e^2 \over h}  
  \Biggl[\, 1 \,
  - 
                \, { 1
                  + 2\, (R-1)^2 
                  \over 3} 
                 \left( {\pi T \over \widetilde{\Delta}}\right)^2
      - 
                \, { 1
                  + 5\, (R-1)^2 
                  \over 4} 
                 \left( {eV \over \widetilde{\Delta} } \right)^2 
 + \cdots 
 \,\Biggr] .              
\label{eq:dI_dV}
\end{equation}
Therefore, the nonlinear $(eV)^2$ term is also scaled 
by the resonance width $\widetilde{\Delta}$, 
and the coefficient is determined by the parameter $R-1$, 
or $\widetilde{U}/(\pi \widetilde{\Delta})$.

\section{Transport theory based on Kubo formalism}
\label{sec:Kubo}

We have discussed in Sec.\ \ref{subsec:current} that 
in the noninteracting case 
the nonequilibrium current can be written 
in a Landauer-type form, as eq.\ (\ref{eq:caro1}). 
The similar expression has been derived 
for interacting electrons in a special case, 
when the couplings between the leads and sample satisfy 
the condition  $\mbox{\boldmath $\Gamma$}^L(\epsilon) 
= \lambda\,\mbox{\boldmath $\Gamma$}^R(\epsilon)$ in 
the notation used in Ref.\ \cite{MW}.
For this condition to be held, 
the interacting sites must be classified 
into the following two groups:   
one group consists of the sites 
that are connected directly to both of the two leads, 
and other group consists of the sites 
that have no direct links (hopping matrix elements) to the leads.   
This condition restricts the application of eq.\ (\ref{eq:caro1}).
For instance, if there is an interacting site 
that is connected to only one of the two leads, 
the condition is not satisfied. 
Therefore, eq.\ (\ref{eq:caro1}) is not applicable 
to a series of quantum dots as illustrated 
in Fig.\ \ref{fig:lattice}.
Nevertheless, in the linear-response regime, 
the Landauer-type expression 
of the dc conductance, eq.\ (\ref{eq:Landauer}),
can be derived quite generally without 
the condition mentioned above \cite{ao10}.

In Sec.\ \ref{subsec:dc_conductance}, 
based on the Kubo formalism, 
we describe the outline of the derivation of 
eq.\ (\ref{eq:Landauer}) for interacting electrons. 
Our proof uses the analytic properties of the vertex corrections 
following along the \'{E}liashberg theory 
of a transport equation for correlated electrons \cite{Eliashberg,AGD}. 
The many-body transmission probability ${\cal T}(\epsilon)$ 
is given by eq.\ (\ref{eq:T_eff2}), 
and it is written in terms of a three-point correlation function. 
At zero temperature,
 the imaginary part 
of the self-energy due to the interaction and the vertex  corrections 
for the current become zero at the Fermi energy $\epsilon=0$. 
Due to this property, 
the transmission probability at $T=0$ is determined by 
the single-particle Green's functions as shown 
in eq.\ (\ref{eq:Teff_at_ground}). 
In Sec.\ \ref{subsec:Ward_general},  
the current conservation law for the correlation functions 
is described with the generalized Ward identity, 
which expresses the relation between the self-energy and current vertex. 
In Sec.\ \ref{subsec:Lehmann}, 
we provide the Lehmann representation of the three-point functions 
to carry out the analytic continuation formally.
It can be also used for nonperturbative calculations 
of ${\cal T}(\epsilon)$. 
We apply this formulation to a finite Hubbard chain 
in Sec.\ \ref{subsec:Hubbard}, 
and show an example of the transmission probability ${\cal T}(\epsilon)$ 
for interacting electrons.

\subsection{Many-body transmission coefficient  ${\cal T}(\epsilon)$}
\label{subsec:dc_conductance}

We now consider the  Hamiltonian ${\cal H}_{\rm tot}^{\rm eq}$ defined 
in eq.\ (\ref{eq:Heq}) again, which is also illustrated in 
Fig.\ \ref{fig:single}.
The dc conductance $g$ can be determined in the Kubo formalism, 
and it corresponds to the $\omega$-linear imaginary part of 
a current-current correlation function  
 $K_{\alpha\alpha'}(\omega+i 0^+)$;
\begin{eqnarray}
&& g  \,=\,  e^2 \, \lim_{\omega \to 0}
     { K_{\alpha\alpha'}(\omega+i 0^+) 
      - K_{\alpha\alpha'}(i 0^+) \over i \, \omega } 
\label{eq:Kubo} \;, 
\\
&& K_{\alpha\alpha'}(i \nu_l) \, = \,   \int_0^{\beta} \! d \tau 
\left\langle T_{\tau}\, J_{\alpha}(\tau) J_{\alpha'}(0) \right\rangle 
      \, e^{i \, \nu_l \tau}, 
\end{eqnarray}
where $\alpha = L$ or $R$.
The retarded correlation can be calculated 
via the analytic continuation 
$K_{\alpha\alpha'}(\omega+i 0^+) \equiv  
\left. K_{\alpha\alpha'}(i \nu_l)\right|_{i \nu_l 
\to \omega + i 0^+}$, where
$\nu_l = 2\pi l/\beta$ is the Matsubara frequency. 
The current operator $J_{\alpha}$ is defined by 
\begin{eqnarray}
 J_L  &=& i \,     
\sum_{\sigma}
                  v_{L}^{\phantom{\dagger}} 
           \left(\, 
            c^{\dagger}_{1 \sigma} 
            c^{\phantom{\dagger}}_{0 \sigma}  
          - c^{\dagger}_{0 \sigma} 
            c^{\phantom{\dagger}}_{1 \sigma}
\, \right) 
\label{eq:J_L}
,\\
 J_R  &=& i \,     
\sum_{\sigma}
         v_{R}^{\phantom{\dagger}}
  \left(\, 
           c^{\dagger}_{N+1 \sigma} 
           c^{\phantom{\dagger}}_{N  \sigma} 
        -  c^{\dagger}_{N \sigma} 
           c^{\phantom{\dagger}}_{N+1  \sigma} 
\, \right) 
\label{eq:J_R} .
\end{eqnarray}
Here $J_L$ is the current flowing into the sample 
from the left lead,  and 
$J_R$ is the current flowing out to the right lead from the sample.
These currents and total charge in the sample $\rho_C^{\phantom{0}}$ 
satisfy the equation of continuity  
\begin{eqnarray}
&& \rho_C^{\phantom{0}} \, = \,     
\sum_{j\in C, \sigma} 
 c^{\dagger}_{j \sigma} c^{\phantom{\dagger}}_{j \sigma} \;,\\
&& {\partial \rho_C^{\phantom{0}} \over \partial t} + J_R - J_L \,=\, 0 \;.
\rule{0cm}{0.5cm}
\end{eqnarray}
Owing to this property, 
the dc conductance  $g$ defined in eq.\ (\ref{eq:Kubo}) does not 
depend on the choice of $\alpha$ and $\alpha'$ \cite{FisherLee}.
Note that  $K_{\alpha'\alpha}(z)  =  K_{\alpha\alpha'}(z)$ 
owing to the time-reversal symmetry 
of ${\cal H}$.

To calculate the $\omega$-linear imaginary part 
of $K_{\alpha\alpha'}(\omega+i 0^+)$,
we introduce the three-point correlation functions of 
the charge and currents,
\begin{eqnarray} 
&& \Phi_{C;jj'}(\tau; \tau_1, \tau_2) 
\ = \     \left \langle  T_{\tau} \, \delta\rho_C^{\phantom{0}}(\tau)\,
 c^{\phantom{\dagger}}_{j \sigma} (\tau_1) \, 
 c^{\dagger}_{j' \sigma} (\tau_2)                      
 \right \rangle ,
\label{eq:Phi_C}
\\
&& \Phi_{L;jj'}(\tau; \tau_1, \tau_2) 
\ = \     \left \langle T_{\tau} \, J_L(\tau)\,
 c^{\phantom{\dagger}}_{j \sigma} (\tau_1) \, 
 c^{\dagger}_{j' \sigma} (\tau_2)                      
 \right \rangle ,
\label{eq:Phi_L}
\\
&& \Phi_{R;jj'}(\tau; \tau_1, \tau_2) 
\ =\     \left \langle  T_{\tau} \, J_R(\tau)\,
 c^{\phantom{\dagger}}_{j \sigma} (\tau_1) \, 
 c^{\dagger}_{j' \sigma} (\tau_2) 
     \right \rangle ,
\label{eq:Phi_R}
\end{eqnarray} 
where 
$\delta \rho_C^{\phantom{0}} \equiv \rho_C^{\phantom{0}} 
 - \langle \rho_C^{\phantom{0}} \rangle$.
These three functions can be expressed as functions 
of two Matsubara frequencies $i\nu$ and $i \varepsilon$,
\begin{equation}
\Phi_{\gamma:jj'}(\tau; \tau_1, \tau_2) 
\, = \, {1 \over \beta^2} 
\sum_{i \varepsilon , i \nu} 
\Phi_{\gamma:jj'}
(i \varepsilon , i \varepsilon  + i \nu)
\, e^{-i \,\varepsilon (\tau_1 - \tau)} 
\, e^{-i \,(\varepsilon + \nu) (\tau - \tau_2)} 
\;,
\label{eq:Matsubara_Fourier}
\end{equation}
for $\gamma = C, L, R$. 
We mainly consider 
the electrons in the central region assuming $jj' \in C$. 
In the right-hand side of eqs.\ (\ref{eq:Phi_L}) and (\ref{eq:Phi_R}), 
there still exist the creation and annihilation operators 
with respect to the leads at $0$ and $N+1$ 
in the operators $J_L$ and $J_R$. 
The correlation functions that include these two sites 
as one of the external points can be related to 
those defined with respect to the adjacent sites $1$ and $N$,   
by using the properties of the Green's function at the two interfaces:
\begin{eqnarray}
  \left \{
  \begin{array}{ll} 
   G_{0, j}(z) \phantom{_{+1}}
 \, =\,  -
  \mbox{\sl g}_L(z)\, v_{L}^{\phantom{\dagger}} G_{1, j}(z)
   &\quad \mbox{for}\quad  1\leq j \leq N+1
   \\
   G_{j, N+1}(z) 
    \, =\,  - 
     G_{j, N}(z)\, v_{R}^{\phantom{\dagger}}\,\mbox{\sl g}_R(z)
      &\quad \mbox{for}\quad  0 \leq j \leq N
      \rule{0cm}{0.5cm}
      \\
 \end{array}
  \right.
   \;\;.
\label{eq:rec_lead} 
\end{eqnarray}
Here $\mbox{\sl g}_L(z)$  and $\mbox{\sl g}_R(z)$ are 
the local Green's functions at the interfaces 
of the isolated leads, $0$ and $N+1$, respectively.
Using these properties, the three-point correlation functions 
for $jj' \in C$ can be expressed as
\begin{eqnarray}
&&
\!\!\!\!\!\!\!\!\!\!\!\!\!\!\!\!\!\!\!\!
  \Phi_{\gamma;jj'}(i \varepsilon , i \varepsilon  + i \nu)
 \, = \sum_{j_4j_1 \in C}
  G_{jj_4}(i \varepsilon )\, 
\Lambda_{\gamma;j_4j_1}(i \varepsilon , i \varepsilon  + i \nu)
  \, G_{j_1j'}(i \varepsilon  + i \nu) 
  \;,
  \label{eq:Lambda}
\end{eqnarray}
where $\Lambda_{\gamma;j_4j_1}$ includes all the vertex corrections.
The corresponding bare current vertices 
are given by 
\begin{eqnarray} 
&& 
\Lambda_{C;j_4j_1}^{(0)}(i \varepsilon, i \varepsilon  + i \nu) \,=\, 
\delta_{j_4,j_1}
\;,
\label{eq:Lambda_C_bare}
\\
&& 
 \,\Lambda_{L;j_4j_1}^{(0)}(i \varepsilon, i \varepsilon  + i \nu) \,=\,
  \lambda_L(i \varepsilon, i \varepsilon  + i \nu) 
\, \delta_{1,j_4} \, \delta_{1,j_1}
\label{eq:Lambda_L_bare}
\;,
\\
&& 
 \,\Lambda_{R;j_4j_1}^{(0)}(i \varepsilon, i \varepsilon  + i \nu) \,=\,
\lambda_R(i \varepsilon, i \varepsilon  + i \nu) 
\, \delta_{N,j_4} \, \delta_{N,j_1}
\;, 
\label{eq:Lambda_R_bare}
\end{eqnarray} 
with  
\begin{eqnarray}
&& \lambda_L(i \varepsilon , i \varepsilon  + i \nu)
\,=\, -i \, v_L^{2} 
\left[\, \mbox{\sl g}_L(i \varepsilon  + i \nu) 
- \mbox{\sl g}_L(i \varepsilon )\,\right]
\;,
\label{eq:lambda_L0} 
\\
&& \lambda_R(i \varepsilon , i \varepsilon  + i \nu)
\,=\, \ i \, v_R^{2} 
\left[\, \mbox{\sl g}_R(i \varepsilon  + i \nu) 
- \mbox{\sl g}_R(i \varepsilon )\,\right]
\;,
\label{eq:lambda_R0} 
 \end{eqnarray}

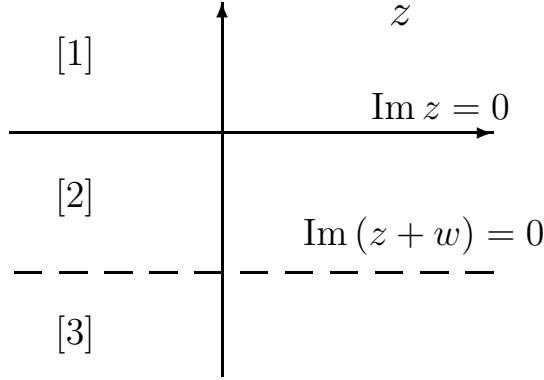
\begin{figure}[tb]
\setlength{\unitlength}{0.6mm}
\begin{center}
\begin{picture}(108,85)(-54,-57) 
\thicklines

\put(-50,0){\vector(1,0){107}}
\put(-3,-54){\vector(0,1){83}}
\multiput(-49,-31)(10,0){11}{\line(1,0){6}}

\put(30,3){\makebox(0,0)[bl]{\large $\mbox{Im}\,z=0$}}
\put(15,-27){\makebox(0,0)[bl]{\large $\mbox{Im}\,(z+w)=0$}}

\put(34,24){\makebox(0,0)[bl]{\Large $z$}}

\put(-40,13){\makebox(0,0)[bl]{\large $[1]$}}
\put(-40,-18.5){\makebox(0,0)[bl]{\large $[2]$}}
\put(-40,-49){\makebox(0,0)[bl]{\large $[3]$}}

\end{picture}
\end{center}
\caption{Three analytic regions of 
$\Phi_{R;11}(z, z+w)$.
}
\label{fig:current_analytic}
\end{figure}

We now calculate the $\omega$-linear part of $K_{\alpha \alpha'}(z)$ 
taking $\alpha$ and $\alpha'$ to be $R$ and $L$, respectively.
Using the three-point correlation functions,  
the current-current correlation function $K_{RL}(i \nu)$ can 
be expressed as
\begin{eqnarray}
K_{RL}(i \nu) \,=                
 \ {1 \over \beta}
\sum_{i \varepsilon } 
\sum_{\sigma} 
\lambda_L(i \varepsilon , i \varepsilon  + i \nu)\,
\, \Phi_{R;11}(i \varepsilon , i \varepsilon  + i \nu)
\;. 
\label{eq:K_nu_new}
\end{eqnarray}
Paying attention to the analytic properties of the 
Green's functions, the summation over the Matsubara frequency 
can be rewritten in a contour-integral form.
Then, by carrying out the analytic continuation 
$i \nu \to \omega + i 0^+$, we obtain
\begin{eqnarray}
&&
\!\!\!\!\!\!\!\!\!\!\!\!\!\!\!\!
K_{RL}(\omega + i 0^+ ) 
\nonumber \\
&&
\quad =   
\,\sum_{\sigma} 
\biggl \{ 
\, - \int_{-\infty}^{\infty} {d \epsilon \over 2 \pi i }\ 
f(\epsilon)
\ \lambda_L^{[1]}(\epsilon,\epsilon + \omega) 
\, \Phi_{R;11}^{[1]}(\epsilon,\epsilon + \omega) 
\nonumber \\
& & \qquad \qquad  - \int_{-\infty}^{\infty} {d \epsilon \over 2 \pi i } \,
\Bigl[\, f(\epsilon+\omega) - f(\epsilon) \,\Bigr]
\ \lambda_L^{[2]}(\epsilon,\epsilon + \omega)
\, \Phi_{R;11}^{[2]}(\epsilon,\epsilon + \omega) 
\nonumber \\
& & \qquad \qquad + \int_{-\infty}^{\infty} {d \epsilon \over 2 \pi i }\ 
f(\epsilon+\omega)
\ \lambda_L^{[3]}(\epsilon,\epsilon + \omega) 
\, \Phi_{R;11}^{[3]}(\epsilon,\epsilon + \omega) 
\, \biggr \} ,
\label{eq:K_ret} 
\end{eqnarray}
where $f(\epsilon) = (e^{\beta \epsilon}+1)^{-1}$.
The superscript with the bracket, 
that is, $[k]$ for $k = 1, 2, 3$,  is introduced 
specifies the three analytic region of
$\Phi_{R;11}(z,z + w)$ and $\lambda_L(z,z+w)$  
in the complex $z$-plane. These regions 
are separated by the two lines, 
$\mbox{Im}\, (z)=0$ and $\mbox{Im}\, (z+w)=0$,  
as shown in Fig \ref{fig:current_analytic}.
In each of the three regions,
 $\Phi_{R;11}(z,z+w)$ corresponds to the analytic function given by   
\begin{equation}
\left\{
\begin{array}{l}
\Phi_{R;11}^{[1]}(\epsilon, \epsilon+\omega) \ = \ 
\Phi_{R;11}(\epsilon +i0^+, \epsilon + \omega +i 0^+)
\\
\Phi_{R;11}^{[2]}(\epsilon, \epsilon+\omega) \ = \ 
\Phi_{R;11}(\epsilon -i 0^+, \epsilon + \omega + i0^+)
\rule{0cm}{0.55cm}
\\
\Phi_{R;11}^{[3]}(\epsilon, \epsilon+\omega) \ = \ 
\Phi_{R;11}(\epsilon -i 0^+, \epsilon + \omega - i 0^+)
\rule{0cm}{0.55cm}
\end{array} \right. 
\label{eq:analytic_continuation}
\;.
\end{equation}
These analytic properties can be clarified explicitly  
in the Lehmann representation, eq.\ (\ref{eq:Lehmann_1}), 
provided in the next subsection. Similarly,
the analytic continuation of 
the bare current 
vertex $\lambda_{\alpha}(i \varepsilon , i \varepsilon  + i \nu)$ 
for $\alpha =L, R$  is given by 
\begin{equation}
 \left\{ 
   \begin{array}{l}
     \lambda_{\alpha}^{[1]}(\epsilon, \epsilon +\omega) \ = \ 
  i s_{\alpha}\, 
 v_{\alpha}^2\left[\, \mbox{\sl g}_{\alpha}^+(\epsilon + \omega) - 
           \mbox{\sl g}_{\alpha}^+(\epsilon)  \,\right]
            \\
     \lambda_{\alpha}^{[2]}(\epsilon, \epsilon +\omega) \ = \ 
   i s_{\alpha}\,
   v_{\alpha}^2\left[\, \mbox{\sl g}_{\alpha}^+(\epsilon + \omega) - 
     \mbox{\sl g}_{\alpha}^-(\epsilon) \,\right] 
       \rule{0cm}{0.55cm}
           \\
     \lambda_{\alpha}^{[3]}(\epsilon, \epsilon +\omega) \ = \ 
    i s_{\alpha}\,
    v_{\alpha}^2\left[ \,  \mbox{\sl g}_{\alpha}^-(\epsilon + \omega) - 
       \mbox{\sl g}_{\alpha}^-(\epsilon) \,\right]
         \rule{0cm}{0.55cm}
    \end{array} \right.
 \;,
\end{equation}
where the factor $s_{\alpha}$ is defined such that $s_{L}=-1$ and  $s_{R}=+1$.
In this section, 
we distinguish the retarded and advanced Green's functions 
by the label $+$ and $-$, respectively, in the superscript.  
In the limit of $\omega \to 0$, 
the bare vertices for $k=1$ and $3$ vanish as  
$\lambda_{\alpha}^{[k]}(\epsilon, \epsilon+\omega) \propto \omega$. 
In contrast, for $k=2$, it tends to a finite constant
$\lambda_{\alpha}^{[2]}(\epsilon, \epsilon)
= 2\,s_{\alpha} \Gamma_{\alpha}(\epsilon)$ with 
$\Gamma_{\alpha}(\epsilon) = -\,v_{\alpha}^2\, \mbox{Im} 
\left[ \mbox{\sl g}_{\alpha}^{+}(\epsilon)  \right]$. 
Correspondingly, 
the asymptotic behavior of $\Phi_{\alpha}^{[k]}(\epsilon, \epsilon +\omega)$ 
for small $\omega$ has been investigated 
by using the Lehmann representation of 
a four-point vertex function  \cite{Eliashberg,AGD}, 
and the result is  \cite{ao10}, 
\begin{equation}
    \Phi_{\alpha}^{[k]}(\epsilon, \epsilon +\omega) \propto
 \left\{ 
   \begin{array}{ll}
   \omega
           &\quad \mbox{for} \quad k=1     
            \\
    {\rm finite}
           &\quad \mbox{for} \quad k=2     
           \\
    \,    
     \omega
               &\quad \mbox{for} \quad k=3      
    \end{array} \right.
 \;,
\end{equation}
for $\alpha =L, R$. 
Therefore,  
taking the $\omega \to 0$ limit in eq.\ (\ref{eq:Kubo})
by using eq.\ (\ref{eq:K_ret}) for $K_{RL}(\omega + i 0^+ )$, 
we obtain
\begin{eqnarray}
 && g \,=\, \frac{2e^2}{h} \int d \epsilon\, 
 \left( - {\partial f \over \partial \epsilon} \right) \, 
 {\cal T}(\epsilon) \;,  
\label{eq:Landauer}
\\
 \nonumber \\
&& {\cal T}(\epsilon)  \,=\,     
  2\, \Gamma_L(\epsilon) \, \Phi_{R;11}^{[2]}(\epsilon, \epsilon) 
\;.
\rule{0cm}{0.6cm}
\label{eq:T_eff2}
\end{eqnarray}
Thus, the dc conductance is determined by 
the three-point function for the analytic region $k=2$. 
The analytic continuation is performed 
formally by using the Lehmann representation 
in Sec.\ \ref{subsec:Lehmann}.
The result shows that $\Phi_{R;11}^{[2]}(\epsilon, \epsilon)$ can be 
expressed as a Fourier transform, eq.\ (\ref{eq:Fourier_time}), 
of a real-time retarded product in eq.\ (\ref{eq:time2}).

Specifically, at $T=0$ the conductance is determined by 
the value of the transmission probability at the Fermi $\epsilon=0$,
and it can be written in the form  \cite{ao5,ao6,ao7,ao9},
\begin{equation}
{\cal T}(0)  \,=\,  
4\  \Gamma_L(0) \, G_{1 N}^{-}(0) \, 
\Gamma_R(0) \, G_{N 1}^{+}(0) 
\;.
\label{eq:Teff_at_ground}
\end{equation}
This is due to the property that 
the vertex corrections for the current vanishes
at $T=0$ and $\epsilon=0$, 
as shown in eq.\  (\ref{eq:current_vertex_at_ground})
in Sec.\ \ref{subsec:Ward_general}.
Furthermore, the reflection probability is given by 
\begin{equation}
{\cal R}(0) \,=\,   
\left|1 -2i \Gamma_L(0)  G_{1 1}^{+}(0)\right|^2 
\,=\, \left|1 -2i \Gamma_R(0)  G_{N N}^{+}(0)\right|^2 \;.
\label{eq:Reff}
\end{equation}
The current conservation ${\cal T}(0) + {\cal R}(0) = 1$  
follows from the identity in  eq.\ (\ref{eq:optical}). 
Similarly,  at zero temperature, 
the Friedel sum rule for interacting electrons is given by
\cite{LangerAmbegaokar}, 
\begin{equation}
   \Delta N_{\rm tot}
\,=\, 
{1 \over \pi i}\  
\log [\, \det 
 \mbox{\boldmath $S$}
\,]   \;,
\rule{4cm}{0cm}
\label{eq:Friedel} 
\end{equation}
where the $S$-matrix is defined by
\begin{equation}
\mbox{\boldmath $S$}
 \,=\,  
  \left [ \,
 \matrix { 
 1 -2i\Gamma_L(0)G_{1 1}^{+}(0) 
 & \ \ -2i\Gamma_L(0)G_{1 N}^{+}(0) \ \  \cr
 -2i\Gamma_R(0)G_{N 1 }^{+}(0) \ \ 
 &  \rule{0cm}{0.5cm} \ 1 -2i\Gamma_R(0)G_{N N }^{+}(0) \cr  }
 \, \right ]  
 \;.
 \label{eq:S}
\end{equation}
In eq.\ (\ref{eq:Friedel}),  
$\,\Delta N_{\rm tot}$ 
is the displacement of the total charge defined by  
\begin{eqnarray}
&& 
\!\!\!\!\!\!\!\!\!\!\!\!\!\!\!\!\!\!\!\!\!\!\!
\Delta N_{\rm tot}
 =   
\sum_{i\in C} \sum_{\sigma} 
 \langle c^{\dagger}_{i \sigma} c^{\phantom{\dagger}}_{i \sigma}
\rangle 
\nonumber
\\
&& 
\!\!
+ \sum_{i\in L} \sum_{\sigma} 
  \left[\, 
  \langle c^{\dagger}_{i \sigma} c^{\phantom{\dagger}}_{i \sigma} \rangle
  - 
  \langle c^{\dagger}_{i \sigma} 
  c^{\phantom{\dagger}}_{i \sigma} \rangle_{L}^{\phantom{0}}
    \,\right] 
+ \sum_{i\in R} \sum_{\sigma} 
  \left[\, 
  \langle c^{\dagger}_{i \sigma} c^{\phantom{\dagger}}_{i \sigma} \rangle
  - 
  \langle c^{\dagger}_{i \sigma} 
  c^{\phantom{\dagger}}_{i \sigma} \rangle_{R}^{\phantom{0}}
    \,\right] , 
\label{eq:dn_def}
\end{eqnarray}
where $\langle \cdots \rangle_{L}^{\phantom{0}}$  and 
$\langle \cdots \rangle_{R}^{\phantom{0}}$  
denote the ground-state average of isolated leads
determined by ${\cal H}_{L}$ and ${\cal H}_{R}$, respectively.


\subsection{Current Conservation and Ward identity}
\label{subsec:Ward_general}

The inter-electron interactions generally cause 
the damping of excitations.
Therefore, theoretically, 
the self-energy and vertex corrections
must be treated consistently with 
the approaches that conserve the current. 
In this subsection, we discuss  
the current conservation using a generalized Ward identity.

The generalized Ward identity can be derived from
the equation of continuity in the Matsubara form 
$- \left({\partial /\partial \tau}\right) \delta \rho_C^{\phantom{0}}
+ i \, J_R - i \,J_L = 0$  \cite{Schrieffer},
\begin{eqnarray}
& & 
\!\!\!\!\!\!\!\!\!\!\!\!\!\!\!\!\!\!\!\!\!\!\!\!\!\!\!\!
-\, {\partial \over  \partial \tau }\,  
\Phi_{C;jj'}(\tau; \tau_1, \tau_2) 
\,
+ i \,  
\Phi_{R;jj'}(\tau; \tau_1, \tau_2) 
- i \,
\Phi_{L;jj'}(\tau; \tau_1, \tau_2) 
\nonumber \\ 
&& \qquad \qquad
\ = \ \, \delta (\tau - \tau_2) \, G_{jj'}(\tau_1,\tau) 
        -\, \delta (\tau_1 - \tau) \, G_{jj'}(\tau,\tau_2) 
\;.
\rule{0cm}{0.4cm}
\label{eq:Ward1} 
\end{eqnarray}
It can be expressed by
using a $N\times N$ matrix representation for $jj' \in  C$ 
with the Matsubara frequencies,  
\begin{eqnarray}
&& 
\!\!\!\!\!\!\!\!\!\!\!\!\!\!\!\!\!\!\!\!\!\!\!\!\!\!\!
i \nu \, 
 \mbox{\boldmath $\Phi$}_C
 (i \varepsilon , i \varepsilon  + i \nu) + i \,  
\mbox{\boldmath $\Phi$}_R
(i \varepsilon, i \varepsilon + i \nu)
 - i \, 
\mbox{\boldmath $\Phi$}_L
(i \varepsilon, i \varepsilon + i \nu)
\nonumber \\
 && 
 \qquad \qquad \qquad \qquad \qquad \qquad \qquad  = \ 
 \mbox{\boldmath $G$}(i \varepsilon ) 
 -
 \mbox{\boldmath $G$}(i \varepsilon +i \nu) \;  . 
\rule{0cm}{0.5cm}
  \label{eq:Ward2} 
\end{eqnarray}
Here, $\mbox{\boldmath $G$}(z) = \{G_{jj'}(z)\}$ and  
$\mbox{\boldmath $\Phi$}_{\gamma}(z,z+w)  = \{\Phi_{\gamma;jj'}(z,z+w)\}$.
The matrix version of eq.\ (\ref{eq:Lambda}) is given by 
\begin{equation}
 \mbox{\boldmath $\Phi$}_{\gamma}(z, z + w)
\ = \  \mbox{\boldmath $G$}(z)\, 
\mbox{\boldmath $\Lambda$}_{\gamma}(z, z + w)
\, \mbox{\boldmath $G$}(z + w) \;.
\end{equation}
Thus, the identity can also be expressed
using  $\mbox{\boldmath $\Lambda$}_{\gamma}(z,z+w) 
= \{\Lambda_{\gamma;jj'}(z,z+w)\}$, as
\begin{eqnarray}
&& 
\!\!\!\!\!\!\!\!\!\!\!\!\!\!\!\!\!\!\!\!\!\!\!\!\!\!\!
i \nu \, 
\mbox{\boldmath $\Lambda$}_C
(i \varepsilon , i \varepsilon  + i \nu)
+ i \, 
\mbox{\boldmath $\Lambda$}_R
(i \varepsilon, i \varepsilon + i \nu)
- i \, \mbox{\boldmath $\Lambda$}_L
(i \varepsilon, i \varepsilon + i \nu)
\nonumber \\
&& 
\qquad \qquad \qquad \qquad \qquad  = \ 
\left\{\mbox{\boldmath $G$}(i \varepsilon +i \nu)\right\}^{-1}  
-
\left\{\mbox{\boldmath $G$}(i \varepsilon )\right\}^{-1} \;.
\rule{0cm}{0.5cm}
\label{eq:Ward3} 
\end{eqnarray}
Furthermore, the Dyson equation for the single-particle Green's function 
can be expressed as 
\begin{eqnarray}
\left\{\mbox{\boldmath $G$}(z)\right\}^{-1}  
&=& 
 z \, \mbox{\boldmath $1$} 
 -  \mbox{\boldmath ${\cal H}$}_C^0  
- \mbox{\boldmath ${\cal V}$}_{\rm mix}(z)  
- \mbox{\boldmath $\Sigma$}(z)  \;,    
\label{eq:Dyson_linear}
\\
\mbox{\boldmath ${\cal H}$}_C^0 &=&  
\left [ \,
 \matrix { -t_{11}^C-\mu    & -t_{12}^C      & \cdots  &               \cr
            -t_{21}^C        & -t_{22}^C -\mu &         &               \cr
            \vdots          &               & \ddots  &               \cr 
                            &               &         & -t_{NN}^C -\mu \cr
          }
          \, \right ]  
\;,
\label{eq:matrix_H0}
\rule{0cm}{1.5cm}
\\
\mbox{\boldmath ${\cal V}$}_{\rm mix}(z) &=& 
 \left [ \, \matrix { 
 v_L^{2} \, \mbox{\sl g}_L(z) & 0 & \cdots &0 & 0 \cr
 0           &    0     & \cdots   &  0 & 0     \cr
 \vdots      &  \vdots  & \ddots   &  \vdots & \vdots  \cr 
 0 & 0 & \cdots & 0 & 0 \cr
 0 & 0 & \cdots & 0 & v_R^{2} \, \mbox{\sl g}_R(z) \cr
                    }
 \, \right ]  
\label{eq:V_mix} 
\rule{0cm}{1.6cm}
\;,
\end{eqnarray}
and $\mbox{\boldmath $\Sigma$}(z) = \{\Sigma_{jj'}(z)\}$ is 
the self-energy due to the inter-electron interactions. 
Therefore,  eq.\ (\ref{eq:Ward3}) represents a relation between 
the self-energy and vertex functions,
and this identity must be satisfied in the conserving approaches.
Carrying out the analytic continuation of eq.\ (\ref{eq:Ward3}) 
in the region $k=2$,
$i\varepsilon  + i\nu \to \epsilon +\omega +i 0^+$ and 
$i\varepsilon  \to \epsilon -i 0^+$, and 
then taking the limit of $\omega \to 0$, we obtain
\begin{equation}
\mbox{\boldmath $\Lambda$}_R^{[2]}
(\epsilon,  \epsilon)
- \mbox{\boldmath $\Lambda$}_L^{[2]}
(\epsilon,  \epsilon) \, = \, 
-2\, \mbox{Im}\, \mbox{\boldmath ${\cal V}$}_{\rm mix}^+(\epsilon)
-2\, \mbox{Im}\, \mbox{\boldmath $\Sigma$}^+(\epsilon) 
\label{eq:Ward_Im}
\end{equation}
At $T=0$, $\epsilon=0$, the imaginary part of the self-energy vanishes 
$\mbox{Im}\, \mbox{\boldmath $\Sigma$}^+(0)=0$, 
and then the current vertices become equal to the bare ones,   
\begin{eqnarray}
&&\Lambda_{R;j_4j_1}^{[2]}(0, 0)
 \,=\, \,2\,\Gamma_R(0)\, \delta_{N j_4}  \delta_{N j_1}
 \;,
 \label{eq:current_vertex_at_ground}
 \\
&&\Lambda_{L;j_4j_1}^{[2]}(0, 0)
 \,=\, -\,2\,\Gamma_L(0)\, \delta_{1 j_4}  \delta_{1 j_1}
\;.
      \rule{0cm}{0.6cm}
\end{eqnarray}
Correspondingly, $\Phi_{R;11}^{[2]}(0, 0) =
G_{1N}^-(0) 2\Gamma_R(0) G_{N1}^+(0)$ at $T=0$, 
and then the transmission probability is given by 
eq.\ (\ref{eq:Teff_at_ground}).
Alternatively, at $T=0$, 
the analytic continuation of eq.\ (\ref{eq:Ward1}) 
in region $k=2$ is written in the form
\begin{equation}
\mbox{\boldmath $G$}^+(0) - 
\mbox{\boldmath $G$}^-(0)
\, = \,
\mbox{\boldmath $G$}^+(0) 
 \left[ 
  \mbox{\boldmath ${\cal V}$}_{\rm mix}^+(0)
   -   
  \mbox{\boldmath ${\cal V}$}_{\rm mix}^-(0)  
 \right] 
 \mbox{\boldmath $G$}^-(0) 
 \;,
 \label{eq:optical}
\end{equation}
and the ($1$, $1$) and ($N$,$N$) matrix elements 
represent the optical theorem 
for eqs.\ (\ref{eq:Teff_at_ground}) and (\ref{eq:Reff}).


Particularly, for the single Anderson impurity at $N=1$, 
eq.\ (\ref{eq:Ward2}) becomes a single-component equation 
 with respect to the impurity site. 
In the region $k=2$, it becomes 
$
\Phi_R^{[2]}(\epsilon,  \epsilon)
- \Phi_L^{[2]}(\epsilon,  \epsilon) = 
G^-(\epsilon) - G^+(\epsilon) 
$ in the limit of $\omega\to 0$.
Furthermore, if the mixing terms have a property 
$\Gamma_L(\epsilon) = \lambda \,\Gamma_R(\epsilon)$,
an additional relation $\Phi_L^{[2]}(\epsilon,  \epsilon) = 
- \lambda\, \Phi_R^{[2]}(\epsilon, \epsilon)$ follows. 
Thus, in this case the dc conductance can be written 
in the form \cite{HDW2,MW},
\begin{equation}
g_{\rm single}^{\phantom{0}} \,  
 = \, {2 e^2 \over h} \, 
\int_{-\infty}^{\infty} 
  d \epsilon  
\, \left(- {\partial f \over \partial \epsilon} \right)
 \frac{4\, \Gamma_L \Gamma_R}{\Gamma_R + \Gamma_L} 
 \left[
  - \mbox{Im}\,G^+(\epsilon) \right ]  .
\label{eq:cond_single}
\end{equation}

\subsection{Lehmann representation for ${\cal T}(\epsilon)$}
\label{subsec:Lehmann}

We now show that the transmission 
probability ${\cal T}(\epsilon)$ can be expressed 
in terms of a real-time retarded product in eq.\ (\ref{eq:time2}) 
via the Fourier transform eq.\ (\ref{eq:Fourier_time}). 
It shows a direct link between   
the transmission probability and dynamic correlation functions. 
To prove it, we first of all derive the Lehmann representation for 
$\Phi_{R;11}(i\varepsilon , i\varepsilon +i\nu)$,
 and then carry out the analytical continuation.

Inserting a complete set of the eigenstates, 
${\cal H}|n\rangle 
= E_n |n\rangle$, into eq.\ (\ref{eq:Phi_R}) 
and using eq.\ (\ref{eq:Matsubara_Fourier}), we obtain
\begin{eqnarray}
&& 
\!\!\!\!\!\!\!\!\!\!\!\!\!\!\!\!\!\!\!\!\!\!\!\!\!
\Phi_{R;11}( i \varepsilon , i \varepsilon + i \nu)
 \ = \  
 {1 \over Z} \sum_{lmn} \,
\langle l|c_{1\sigma}^{\dagger}|m\rangle
\langle m|J_R|n\rangle
\langle n|c_{1\sigma}^{\phantom{\dagger}}|l\rangle 
\nonumber\\
&& 
\qquad \qquad 
\ \ 
\times
\Biggl[\; 
{ e^{-\beta E_m} \over 
  (i\varepsilon +i\nu + E_m-E_l)(i\nu +E_m -E_n)} 
\nonumber \\  
&& \qquad \qquad 
\ \ \ \ \ \ \    
-\,{ e^{-\beta E_l} \over 
  (i\varepsilon  + E_n-E_l)
  (i\varepsilon +i\nu +E_m -E_l)} 
\nonumber \\  
&& \qquad \qquad 
\ \ \ \ \  \ \  
-\,{ e^{-\beta E_n} \over 
  (i\nu + E_m-E_n)(i\varepsilon  +E_n -E_l)} 
\; \Biggr] 
\nonumber \\  
&& 
\nonumber \\  
&&\qquad \qquad 
 +     
 {1 \over Z} \sum_{lmn} \,
\langle l|c_{1\sigma}^{\phantom{\dagger}}|n\rangle
\langle n|J_R|m\rangle
\langle m|c_{1\sigma}^{\dagger}|l\rangle 
\nonumber\\
&&\qquad \qquad 
 \ \ \ \ \times
\Biggl[\; 
{ e^{-\beta E_n} \over 
  (i\varepsilon  + E_l-E_n)(i\nu +E_n -E_m)} 
\nonumber \\  
&&\qquad \qquad 
\ \ \ \ \ \ \ \  
+\,{ e^{-\beta E_l} \over 
  (i\varepsilon  + E_l-E_n)
  (i\varepsilon +i\nu +E_l -E_m)} 
\nonumber \\  
&&\qquad \qquad 
\ \ \ \ \ \ \ \ 
-\,{ e^{-\beta E_m} \over 
  (i\varepsilon  +i\nu+ E_l-E_m)(i\nu +E_n -E_m)} 
\; \Biggr] , 
\label{eq:Lehmann_1}
\end{eqnarray}
where $Z = \mbox{Tr}\, e^{-\beta {\cal H}}$. 
From eq.\ (\ref{eq:Lehmann_1}),  
the analytic continuation to obtain
 $\Phi_{R;11}^{[k]}(\epsilon, \epsilon+\omega)$ for $k=1,\,2,\,3$ 
can be carried out by replacing the imaginary frequencies 
$i\varepsilon$ and $i\nu$ by  
the real ones $\epsilon$ and $\omega$, respectively, 
with the infinitesimal imaginary parts shown 
in eq.\ (\ref{eq:analytic_continuation}). 
Then the same expressions for 
$\Phi_{R;11}^{[k]}(\epsilon, \epsilon+\omega)$ 
can be derived from the real-time functions  
\begin{eqnarray}
\Phi_{R;11}^{[1]}(t ; t_1, t_2) 
 &=& 
\theta(t-t_1)\,
\theta(t_1-t_2)\,
\left \langle 
\left[\,
\left\{ c_{1\sigma}^{\phantom{\dagger}}(t_1) 
\,,\, c_{1\sigma}^{\dagger}(t_2)\right\} 
\,, J_R(t) \right]\,
 \right \rangle  
 \nonumber \\
&&   + \, \theta(t_1-t)\,
\theta(t-t_2)\,
\left \langle 
\left\{ c_{1\sigma}^{\phantom{\dagger}}(t_1) \,,\, 
\left[c_{1\sigma}^{\dagger}(t_2)\,,\, J_R(t) \right] \right\}
\, \right \rangle  ,
      \rule{0cm}{0.6cm}
 \nonumber \\
\label{eq:time1}
\\
\Phi_{R;11}^{[2]}(t ; t_1, t_2) 
&=& 
\ \theta(t-t_1)\,
\theta(t_1-t_2)\,
\left \langle 
\left\{ c_{1\sigma}^{\dagger}(t_2) \,,\, 
\left[c_{1\sigma}^{\phantom{\dagger}}(t_1)\,,\, J_R(t) \right] \right\}
\, \right \rangle  
\nonumber \\
&&    - \, \theta(t-t_2)\,
\theta(t_2-t_1)\,
\left \langle 
\left\{ c_{1\sigma}^{\phantom{\dagger}}(t_1) \,,\, 
\left[c_{1\sigma}^{\dagger}(t_2)\,,\, J_R(t) \right] \right\}
\, \right \rangle ,  
      \rule{0cm}{0.6cm}
 \nonumber \\
\label{eq:time2}
\\
\Phi_{R;11}^{[3]}(t ; t_1, t_2) 
 &=& 
-\, \theta(t-t_2)\,
\theta(t_2-t_1)\,
\left \langle 
\left[\,
\left\{ c_{1\sigma}^{\phantom{\dagger}}(t_1) 
\,,\, c_{1\sigma}^{\dagger}(t_2)\right\} 
\,, J_R(t) \right]\,
 \right \rangle  
 \nonumber \\
&&    - \, \theta(t_2-t)\,
\theta(t-t_1)\,
\left \langle 
\left\{ c_{1\sigma}^{\dagger}(t_2) \,,\, 
\left[c_{1\sigma}^{\phantom{\dagger}}(t_1)\,,\, J_R(t) \right] \right\}
\, \right \rangle  , 
      \rule{0cm}{0.6cm}
\nonumber \\
\label{eq:time3}
\end{eqnarray}
where $J_R(t) \equiv e^{i {\cal H} t} J_R e^{- i {\cal H}t}$,
and $\theta(t)$ is the step function. 
The commutators are defined by $[A,B] \equiv AB-BA$, 
and $\{A,B\}\equiv AB+BA$, as usual. 
The Fourier transform into the real frequencies is given by
\begin{eqnarray}
&& 
\!\!\!\!\!\!\!\!\!\!\!\!\!\!\!\!\!\!\!\!\!\!\!\!
\int_{-\infty}^{\infty} 
d t\, d t_1\, d t_2\, 
e^{i\omega t} 
e^{i\epsilon  t_1} 
e^{-i\epsilon' t_2}\, 
\Phi_{R;11}^{[k]}(t ; t_1, t_2) 
 \nonumber \\
 && \qquad \qquad \qquad  \qquad
\  = \  
     2\pi\, \delta(\epsilon+\omega-\epsilon')\, 
    \Phi_{R;11}^{[k]}(\epsilon, \epsilon + \omega)  
\;.
\label{eq:Fourier_time}
\end{eqnarray}
For example, a time-ordered function  
\begin{equation}
F(t ; t_1, t_2)  \ = \ 
\theta(t-t_1)\,
\theta(t_1-t_2)
\left \langle J_R(t)\,
 c_{1\sigma}^{\phantom{\dagger}}(t_1) \, c_{1\sigma}^{\dagger}(t_2)       
 \right \rangle 
\end{equation}
is transformed into 
\begin{eqnarray}
 F(\epsilon,\epsilon+\omega) \ = \ 
 {-1 \over Z} \sum_{lmn} \,
{
e^{-\beta E_m} \, 
\langle l|c_{1\sigma}^{\dagger}|m\rangle
\langle m|J_R|n\rangle
\langle n|c_{1\sigma}^{\phantom{\dagger}}|l\rangle 
\over
  (\epsilon + \omega + E_m-E_l +i0^+)(\omega +E_m -E_n+i0^+)
  } 
  \;.
\nonumber \\
\end{eqnarray}

Among the three real-time functions eqs.\ (\ref{eq:time1})--(\ref{eq:time3}),
the function for the region $k=2$, that is, 
 $\Phi_{R;11}^{[2]}(t ; t_1, t_2)$ in eq.\ (\ref{eq:time2}) determines
the transmission probability ${\cal T}(\epsilon) =  2 \Gamma_L(\epsilon) 
\, \Phi_{R;11}^{[2]}(\epsilon, \epsilon)$. 
Because the analytic continuation has already been done, 
the real-time correlation links directly to the transport coefficient. 
This formulation can be used for numerical calculations.

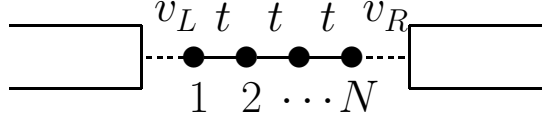
\begin{figure}[t]
\begin{center}
\setlength{\unitlength}{0.7mm}

\begin{picture}(130,20)(-5,0)
\hspace{-0.8cm}
\thicklines

\put(15,4){\line(1,0){25}}
\put(15,16){\line(1,0){25}}
\put(40,4){\line(0,1){12}}

\put(91,4){\line(1,0){25}}
\put(91,16){\line(1,0){25}}
\put(91,4){\line(0,1){12}}

\multiput(41.0,10)(2,0){4}{\line(1,0){1}}
\multiput(82.9,10)(2,0){4}{\line(1,0){1}}

\put(50,10){\circle*{4}} 
\put(60,10){\circle*{4}} 
\put(70,10){\circle*{4}} 
\put(80,10){\circle*{4}} 

\put(52,10){\line(1,0){6}}
\put(62,10){\line(1,0){6}}
\put(72,10){\line(1,0){6}}

\put(54,15){\makebox(0,0)[bl]{\Large $t$}}
\put(64,15){\makebox(0,0)[bl]{\Large $t$}}
\put(74,15){\makebox(0,0)[bl]{\Large $t$}}
\put(42.5,15){\makebox(0,0)[bl]
{\Large $v_L^{\phantom{\dagger}}$}}
\put(82,15){\makebox(0,0)[bl]
{\Large $v_R^{\phantom{\dagger}}$}}

\put(49,-0.5){\makebox(0,0)[bl]{\Large $1$}}
\put(59,-0.5){\makebox(0,0)[bl]{\Large $2$}}
\put(66.5,-0.5){\makebox(0,0)[bl]{\Large $\cdots$}}
\put(77.5,-0.5){\makebox(0,0)[bl]{\Large $N$}}

\end{picture}
\caption{Schematic picture of a finite Hubbard chain.}
\label{fig:lattice}
\end{center}
\end{figure}


\begin{figure}[b]
\leavevmode 
\rule{0.5cm}{0cm}
\begin{minipage}[t]{1\linewidth}
\includegraphics[width=0.2\linewidth]{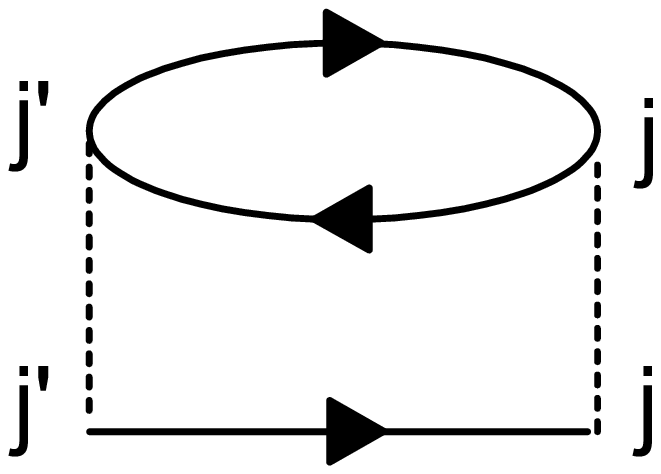}
\label{fig:diagramSelf}
\rule{1.25cm}{0cm}
\includegraphics[width=0.6\linewidth]{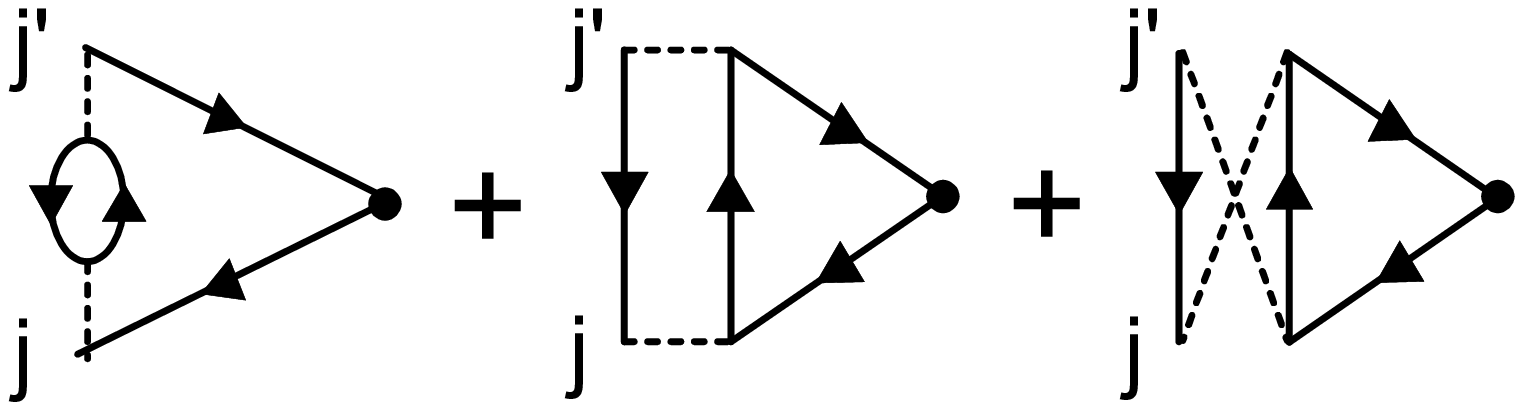}
\label{fig:diagramVertex} \\
{\large \bf \rule{0.9cm}{0cm} (a) \rule{5.5cm}{0cm} (b) }
\rule{0cm}{0.6cm}
\end{minipage}
\caption{The order $U^2$ terms of  
(a) self-energy and (b) vertex corrections.}
\label{fig:order_U2}
\end{figure}

\subsection{Application to a Hubbard chain connected to leads}
\label{subsec:Hubbard}

In this subsection, 
we apply the linear-response formulation 
to a finite Hubbard chain attached to reservoirs,
which can be considered as a model for a series 
of quantum dots or atomic wires of nanometer size.
A schematic picture of the model is shown in Fig.\ \ref{fig:lattice}. 
The Hamiltonian parameters defined in eq.\ (\ref{eq:Heq}) are 
taken as follows.  
We take $t_{ij}^C$ to be the nearest-neighbor hopping $t$, 
and $U_{j_4 j_3; j_2 j_1}$ to be an onsite repulsion $U$. 
Specifically, we consider the electron-hole symmetric case, 
at which $\mu =0$ and $\epsilon_d + U/2 =0$ with  
$\epsilon_d$ being the onsite energy.
We also assume that the two couplings are symmetric 
$\Gamma_L =\Gamma_R$ ($\equiv \Gamma$), 
and the local density of states of the leads is a constant. 

\begin{figure}[t]
\leavevmode 
\begin{center}
\includegraphics[width=\linewidth]{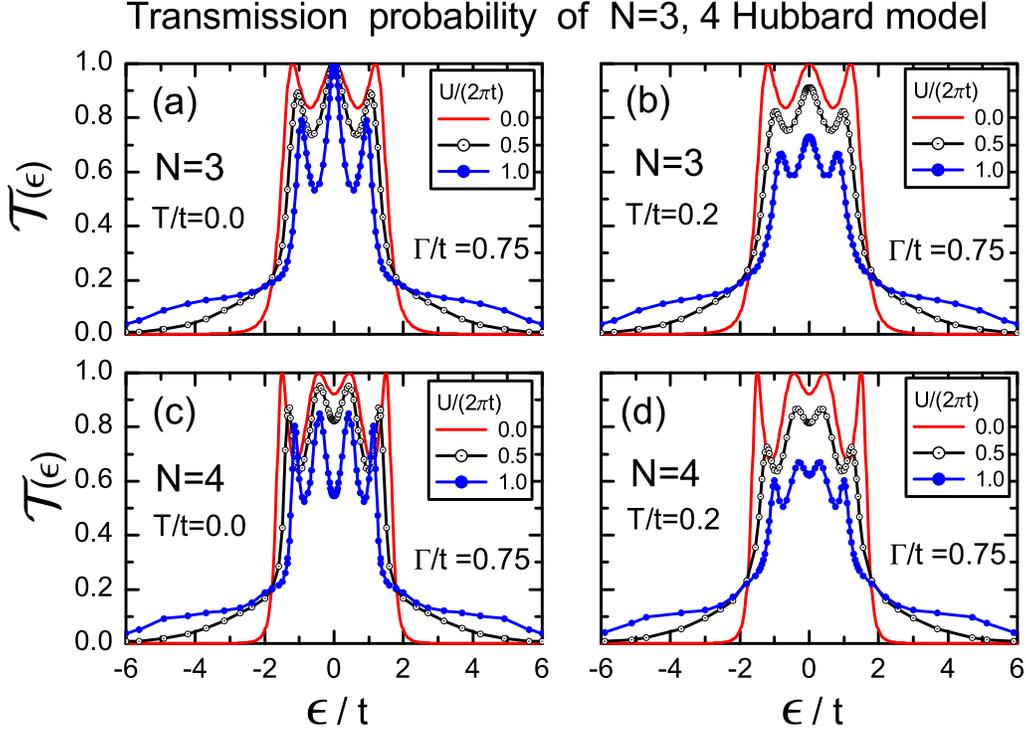}
\end{center}
\caption{
Many-body transmission coefficient for  $N=3$ (upper two panels) 
and $4$ (lower two panels) is plotted vs $\epsilon /t$  
for $\Gamma/t =0.75$  for three values of $U/(2\pi t)$;
(---) $0.0$, (--$\circ$--) $0.5$, and (--$\bullet$--) $1.0$.
The temperature is taken to be
$T/t=0.0$ for (a) and (c),  and $T/t=0.2$ for (b) and (d). 
}
\label{fig:lowT2004}
\end{figure}

To examine the effects of the Coulomb interaction,
we calculate the self-energy and vertex corrections 
up to terms of order $U^2$, the Feynman diagrams for which 
are illustrated in  Fig.\ \ref{fig:order_U2} \cite{ao10}. 
These contributions satisfy 
the generalized Ward identity eq.\ (\ref{eq:Ward_Im})
that corresponds to the current conservation law. 
In Fig.\ \ref{fig:lowT2004}, 
the results of ${\cal T}(\epsilon)$ for 
 $N=3$, $4$ are plotted  
vs $\epsilon/t$ 
for $\Gamma/t = 0.75$ 
for three values of $U/(2\pi t)$;   
(---) $0.0$, (--$\circ$--) $0.5$, and (--$\bullet$--) $1.0$.
The temperature $T/t$ is taken to be    
(a) $0.0$, (b) $0.2$ for $N=3$ in the upper panels,
and  (c) $0.0$, (d) $0.2$ for $N=4$ in the lower panels.

\begin{figure}[t]
\leavevmode 
\begin{center}
\includegraphics[width=0.8\linewidth]{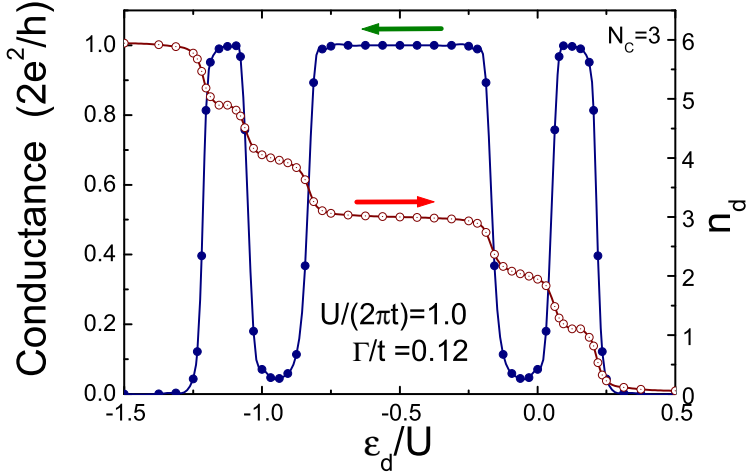}
\end{center}
\caption{
NRG results of the counductance $g$ and 
and the number of electons $n_d$ in triple dots $N=3$ 
as functions of $\epsilon_d$ at $T=0.0$.
}
\label{fig:g3_nrg}
\end{figure}

At low temperatures,  
there are $N$ resonance peaks that have 
 one-to-one correspondence 
to resonant states of the unperturbed system.
In addition to these resonance peaks,   
two broad peaks of atomic character 
appear at $\epsilon \simeq \pm U/2$ for large $U$.
The resonance peaks become sharper with increasing $U$ 
at low temperatures as seen in the panels (a) and (c). 
However, the height of the peaks decreases with increasing $U$.
One exception, which happens for odd $N$,   
is the Kondo resonance at the Fermi level $\epsilon=0$.
At this peak the transmission probability reaches  
the unitary-limit value $1.0$ for any values of $U$,
when the systems have the inversion 
symmetry $\Gamma_L = \Gamma_R$ together with 
the electron-hole symmetry \cite{ao9,ao17}.  
The width of the Kondo 
resonance $T_K$ must decrease with increasing $N$.
For even $N$, the transmission probability   
${\cal T}(\epsilon)$  shows a minimum at $\epsilon=0$.
The characteristic energy scale in this case is  
the width of the valley, which eventually becomes  
the Mott-Hubbard gap in the limit of large $N$.
The high energy profile of ${\cal T}(\epsilon)$ 
at $|\epsilon| \gtrsim 2t$ in the case of $N=3$ 
is similar to that for $N=4$. Namely, the high-energy part 
shows no notable $N$ dependence.
For $U/(2\pi t) =0.5$,  
the upper and lower Hubbard levels at $\epsilon \simeq \pm U/2$ 
exist inside the energy region corresponding to 
the one-dimensional band of the width $2t$. 
The two Hubbard levels got outside of this energy region 
for $U/2 \gtrsim 2t$. 
At finite temperatures, 
the resonance peaks at $|\epsilon| \lesssim 2t$ 
become broad, and the peak height decreases with increasing $T$.
The structures of the resonance peaks vanish 
eventually at higher temperatures, 
and then the even-odd oscillatory behavior disappears \cite{ao10}.

Recently, there has been a numerical progress in this subject.
The conductance for $N=3$ and that of $4$ have been calculated 
away from half-filling as functions of $\epsilon_d$ with numerical 
renormalization group (NRG). 
It has been clarified that the conductance 
shows a typical Kondo behavior as seen in 
Fig.\ \ref{fig:g3_nrg}. Namely,
the plateau of the Unitary limit $g \simeq 2e^2/h$ 
emerges at the gate voltages, $\epsilon_d$, corresponding to 
odd-number occupations of the electrons.
In contrast, the conductance shows  wide minimum 
when the interacting region is occupied by electrons of even numbers 
 \cite{ONH,NO}.

\section{Tomonaga-Luttinger Model}
\label{sec:Tomonaga-Luttinger}

Transport through interacting systems in one dimension 
has been studied extensively  
for quantum wires, organic conductors,
carbon nanotube, etc.  
In this section, we provide a brief introduction to 
a Tomonaga-Luttinger model \cite{Solyom},
to take a quick look at the transport properties 
of a typical interacting system in one dimension.

\subsection{Spin-less fermions in one dimension }

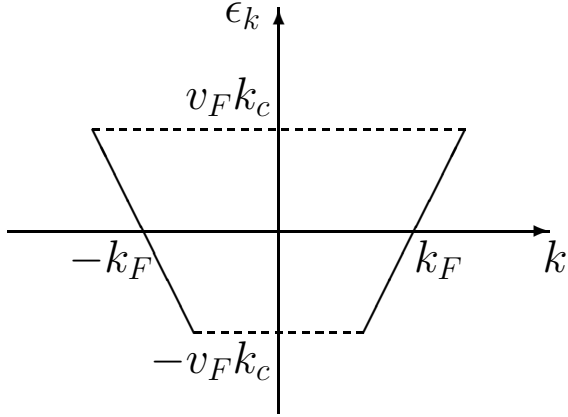
\begin{figure}[t]

\setlength{\unitlength}{0.9mm}
\begin{center}
\begin{picture}(100,65)
\thicklines

\put(10,30){\vector(1,0){80}}
\put(50,3){\vector(0,1){60}}

\put(62.5,15){\line(1,2){15}}
\put(37.5,15){\line(-1,2){15}}

\multiput(37.5,15)(2,0){13}{\line(1,0){1}}
\multiput(22.5,45)(2,0){28}{\line(1,0){1}}

\put(89,24){\makebox(0,0)[bl]{\Large $k$}}
\put(42,60){\makebox(0,0)[bl]{\Large $\epsilon_k$}}

\put(70,23){\makebox(0,0)[bl]{\Large $k_F$}}
\put(19,23){\makebox(0,0)[bl]{\Large $-k_F$}}

 \put(31,47){\makebox(0,0)[bl]{\Large $\phantom{-}v_F k_c$}}
 \put(31,8){\makebox(0,0)[bl]{\Large $-v_F k_c$}}
%


\end{picture}
\end{center}
\label{fig:linear-dispersion}
\caption{Linearized dispersion}
\end{figure}

We start with the spin-less fermions 
described  by the Hamiltonian,
\begin{eqnarray}
& & H_0  \, - \,   \langle H_0 \rangle_0
\, = \,    
   \ \sum_{k} \ (\epsilon_k\, -\,\mu)\,
    \left[\, c^{\dagger}_{k} c^{\phantom{\dagger}}_{k}  
  \,- \, 
  \langle c^{\dagger}_{k} c^{\phantom{\dagger}}_{k} \rangle_0
  \,\right] 
    \;,
\label{eq:H0_1d}
\\
& &  H_I  \, = \,  {1\over 2L} \sum_{qkk'} V_q \,
   c^{\dagger}_{k+q}\,
   c^{\dagger}_{k'-q}\,
   c^{\phantom{\dagger}}_{k'}\,
   c^{\phantom{\dagger}}_{k}
\;.
\label{eq:H_I_1d}
\end{eqnarray}
In eq.\ (\ref{eq:H0_1d}),
the ground-state energy for the noninteracting electrons 
has been subtracted. 
At low energies, the excitations near the Fermi level
play a dominant role, so that $\epsilon_k$ can be 
linearlized at the two Fermi points $k=\pm k_F^{\phantom{0}}$, 
\begin{eqnarray}
&&
 {\cal H}_0 \, = \,    
    \sum_{k} 
  v_F^{\phantom{0}} (k - k_F^{\phantom{0}} ) 
    \left[\,
    a^{\dagger}_{k} a^{\phantom{\dagger}}_{k} 
       -  
   \langle a^{\dagger}_{k} a^{\phantom{\dagger}}_{k} \rangle_0 \,\right] 
  \nonumber   \\
& & \ \qquad
  \, + \, \sum_{k} 
  v_F^{\phantom{0}} (-k - k_F^{\phantom{0}} ) 
    \left[\, b^{\dagger}_{k} b^{\phantom{\dagger}}_{k}  
  - \langle b^{\dagger}_{k} b^{\phantom{\dagger}}_{k} \rangle_0
  \,\right] . 
\label{eq:H0_TL}   
\end{eqnarray}
Here $a_k$ ($b_k$) is the operator 
for the right-moving (left-moving) particles.  
The summation over $k$ in eq.\  (\ref{eq:H0_TL}) 
should be restricted in a range $ |k|-k_F < k_c$
with the cut-off momentum $k_c$ 
of the order a band width $D \sim v_F^{\phantom{0}} k_c^{\phantom{0}}$ 
as illustrated in Fig.\ \ref{fig:linear-dispersion}.
However, we assume that $k_c \to \infty$, 
and will introduce the cut-off for the momentum transfer $q$,
when it is required \cite{Solyom}. 
For low-energy properties, 
the interactions between the electrons near the Fermi level 
is important. Therefore, the interaction Hamiltonian
eq.\ (\ref{eq:H_I_1d}) can be simplified by 
taking only the scattering processes in which 
all the four momentums are close to one of the two Fermi points, that is,
 $k+q \simeq \pm k_F^{\phantom{0}}$, 
$k'-q \simeq \pm k_F^{\phantom{0}}$,
$k' \simeq \pm k_F^{\phantom{0}}$, and $k \simeq \pm k_F^{\phantom{0}}$,
into account;  
\begin{eqnarray}
&& 
\!\!\!\!\!\!\!\!\!\!
H_I   \simeq  {1\over 2L} 
\left(\sum_{q\simeq 0} V_q  
+ \!\! \sum_{q\simeq \pm 2k_F^{\phantom{0}}}  \!\!\! V_{q} \right)
\! \left(\sum_{k\simeq k_F^{\phantom{0}}} 
+ \!\! \sum_{k\simeq -k_F^{\phantom{0}}} \right)
\! \left(\sum_{k'\simeq k_F^{\phantom{0}}} 
+ \!\! \sum_{k'\simeq -k_F^{\phantom{0}}} \right)
\nonumber 
   c^{\dagger}_{k+q}\,
   c^{\dagger}_{k'-q}\,
   c^{\phantom{\dagger}}_{k'}\,
   c^{\phantom{\dagger}}_{k}
\,. \\
&&
\label{eq:H_I_1d_2}
\end{eqnarray}
For the scattering process with small momentum transfer $q \simeq 0$,
there are two types of possibilities for the initial momentums 
 $k \simeq k'$ and $k \simeq -k'$. 
In contrast, in the case of the back scattering
$q \simeq \pm 2k_F^{\phantom{0}}$, 
the incident momentums must have the opposite sign $k \simeq -k'$.
These scattering processes can be described by a
simplified Hamiltonian,
\begin{eqnarray}
& &
\!\!\!\!\!\!\!\!\!\!\!\!\!
H_I \ \Rightarrow  \ {\cal H}_4 \ + \ {\cal H}_2 \ + \ {\cal H}_1  
\;,
\\
& & 
\!\!\!\!\!\!\!\!\!\!\!\!\!
{\cal H}_4
\ = \ 
{g_4 \over 2L} 
\sum_{qkk'}
   a^{\dagger}_{k+q}\,
   a^{\dagger}_{k'-q}\,
   a^{\phantom{\dagger}}_{k'}\,
   a^{\phantom{\dagger}}_{k}
\ + \ 
{g_4 \over 2L} 
\sum_{qkk'}
   b^{\dagger}_{k+q}\,
   b^{\dagger}_{k'-q}\,
   b^{\phantom{\dagger}}_{k'}\,
   b^{\phantom{\dagger}}_{k}
\label{eq:H4_TL}
\;, 
\rule{2cm}{0cm}
\rule{0cm}{0.8cm}
\\
& &
\!\!\!\!\!\!\!\!\!\!\!\!\!
{\cal H}_2
\ = \ 
{g_2 \over 2L} 
\sum_{qkk'}
   a^{\dagger}_{k+q}\,
   b^{\dagger}_{k'-q}\,
   b^{\phantom{\dagger}}_{k'}\,
   a^{\phantom{\dagger}}_{k}
\ + \ 
{g_2 \over 2L} 
\sum_{qkk'}
   b^{\dagger}_{k+q}\,
   a^{\dagger}_{k'-q}\,
   a^{\phantom{\dagger}}_{k'}\,
   b^{\phantom{\dagger}}_{k}
\label{eq:H2_TL}
\;, 
\\
& &
\!\!\!\!\!\!\!\!\!\!\!\!\!
{\cal H}_1
\ = \
{g_1 \over 2L} 
\sum_{q'kk'}\!\!
   b^{\dagger}_{k+q'-2k_F^{\phantom{0}}}\,
   a^{\dagger}_{k'-q' +2k_F^{\phantom{0}}}\,
   b^{\phantom{\dagger}}_{k'}\,
   a^{\phantom{\dagger}}_{k}
\, + \, 
{g_1 \over 2L} 
\sum_{q'kk'}\!\!
   a^{\dagger}_{k+q'+2k_F^{\phantom{0}}}\,
   b^{\dagger}_{k'-q'-2k_F^{\phantom{0}}}\,
   a^{\phantom{\dagger}}_{k'}\,
   b^{\phantom{\dagger}}_{k}
\;. 
\nonumber \\
\label{eq:H1_TL}
\end{eqnarray}
In eq.\ (\ref{eq:H1_TL}), $q'$ is a small momentum 
defined  such that $q=q'\pm 2k_F^{\phantom{0}}$.
The coupling constants should be taken as $g_2 \simeq g_4 \simeq V_0$,
and $g_1 \simeq V_{2k_F}$. However, 
in the following we treat these three constants 
to be independent parameters.
The momentum-transfer cut-off 
is introduced for the summation over $q$ and $q'$.
The Tomonaga-Luttinger model is defined by
\begin{equation}  
{\cal H}_{\rm TL} \, = \, {\cal H}_{0} \, + \, 
{\cal H}_{4} \, + \, {\cal H}_{2} \;.
\label{eq:H_TL}
\end{equation}  
Note that for the spin-less model there should be no distinction 
between  ${\cal H}_2$ and ${\cal H}_1$ \cite{Solyom}.  

The interactions ${\cal H}_{4}$ and ${\cal H}_{2}$ can be 
expressed in terms of the density 
operators $\rho_1(p)$ and  $\rho_2(p)$ defined by
\begin{eqnarray}
\rho_1^{\phantom{0}}(p) &=& \sum_k \left[\,
   a^{\dagger}_{k-p}\,a^{\phantom{\dagger}}_{k}\,
\, - \, \delta_{p,0}\,
\langle a^{\dagger}_{k} a^{\phantom{\dagger}}_{k} \rangle_0 \,\right] 
\label{eq:rho_1_TL}
\;,
\\
\rho_2^{\phantom{0}}(p) &=& \sum_k \left[\,
   b^{\dagger}_{k-p}\,b^{\phantom{\dagger}}_{k}\,
\, - \, \delta_{p,0}\,
\langle b^{\dagger}_{k} b^{\phantom{\dagger}}_{k} \rangle_0 \,\right] 
\;.
\label{eq:rho_2_TL}
\end{eqnarray}
Here 
$\langle a^{\dagger}_{k} a^{\phantom{\dagger}}_{k} \rangle_0 = 
\theta(k_F^{\phantom{0}} -k)$ and
$\langle b^{\dagger}_{k} b^{\phantom{\dagger}}_{k} \rangle_0 = 
\theta(k_F^{\phantom{0}}+k)$, and
these terms are required to define  
the deviation from the noninteracting value 
without an ambiguity caused by 
the occupation of the negative energy states.
Equations (\ref{eq:H4_TL}) and (\ref{eq:H2_TL}) can be 
rewritten in the following forms apart from 
 a renormalization of the chemical potential 
that can be absorbed in to $k_F$,
\begin{eqnarray}
& & 
{\cal H}_4
\ = \ 
{g_4 \over L}\, 
\sum_{p>0}\,
\rho_1^{\phantom{0}}(-p)\, 
\rho_1^{\phantom{0}}(p) 
\ + \ 
{g_4 \over L} \,
\sum_{p>0}\,
\rho_2^{\phantom{0}}(-p)\, 
\rho_2^{\phantom{0}}(p) 
\label{eq:H4_TL2}
\;, 
\rule{2cm}{0cm}
\rule{0cm}{0.8cm}
\\
& &
{\cal H}_2
\ = \ 
{g_2 \over L} \,
\sum_{p>0}\,
\rho_1^{\phantom{0}}(-p)\, 
\rho_2^{\phantom{0}}(p) 
\ + \ 
{g_2 \over L} \,
\sum_{p>0}\,
\rho_2^{\phantom{0}}(-p)\, 
\rho_1^{\phantom{0}}(p) \;.
\label{eq:H2_TL2}
\end{eqnarray}

These two density operators satisfy 
the commutation relations 
\begin{equation}
\left[\,\rho_1^{\phantom{0}}(p) \,,\,\rho_1^{\phantom{0}}(-p')\,\right]
\ = \ {Lp \over 2 \pi} \, \delta_{pp'}
\;, \qquad 
\left[\,\rho_2^{\phantom{0}}(-p) \,,\,\rho_2^{\phantom{0}}(p')\,\right]
\ = \ {Lp \over 2 \pi} \, \delta_{pp'}
\;.
\end{equation}
One notable feature is that these two commutation relations 
are equivalent to those of the bose operators, 
\begin{eqnarray}
& &
C_p^{\phantom{\dagger}} \ = \ \sqrt{2\pi \over Lp} \ \rho_1^{\phantom{0}}(p)
\;, \qquad \qquad
C_p^{\dagger} \ = \ \sqrt{2\pi \over Lp} \ \rho_1^{\phantom{0}}(-p)
\;,
\\
& &
C_{-p}^{\phantom{\dagger}} \ = \ 
\sqrt{2\pi \over Lp} \ \rho_2^{\phantom{0}}(-p)
\;, \qquad 
C_{-p}^{\dagger} \ = \ \sqrt{2\pi \over Lp} \ \rho_2^{\phantom{0}}(p)
\;,
\rule{0cm}{0.8cm}
 \\
 & &
 \left[\,C_p^{\phantom{\dagger}}  \,,\,C_{p'}^{\dagger} \,\right]
 \ = \  \delta_{pp'}
 \;, \qquad \quad
 \left[\,C_{-p}^{\phantom{\dagger}}  \,,\,C_{-p'}^{\dagger} \,\right]
 \ = \  \delta_{pp'} \;,
 \rule{0cm}{0.8cm}
\rule{2cm}{0cm}
\end{eqnarray}
where $p>0$.
The commutation relation of the density operators 
and  Hamiltonian can be calculated by using 
eqs.\ (\ref{eq:H0_TL}) and (\ref{eq:rho_1_TL})--(\ref{eq:H2_TL2}) as
\begin{eqnarray}
&&
\!\!\!\!\!\!\!\!\!\!\!\!\!\!
\left[\,\rho_1^{\phantom{0}}(p) \,,\,{\cal H}_0\,\right]
\,=\,  v_F^{\phantom{0}} \,p   \,\rho_1^{\phantom{0}}(p)
\;, \quad \quad
\left[\,\rho_2^{\phantom{,0}}(-p) \,,\,{\cal H}_0\,\right]
\ = \,  v_F^{\phantom{0}} \,p   \,\rho_2^{\phantom{0}}(-p) \;,
\label{eq:commutator_0}
\\
&&
\!\!\!\!\!\!\!\!\!\!\!\!\!\!
\left[\,\rho_1^{\phantom{0}}(p) \,,\,{\cal H}_4\,\right]
\,=\,  \widetilde{g}_4 v_F^{\phantom{0}}\,p   \,\rho_1^{\phantom{0}}(p)
\;, \quad 
\left[\,\rho_2^{\phantom{0}}(-p) \,,\,{\cal H}_4\,\right]
\, = \,   \widetilde{g}_4 v_F^{\phantom{0}} \,p   \,\rho_2^{\phantom{0}}(-p)
\;,
\rule{0cm}{0.5cm}
\label{eq:commutator_4}
\\
&&
\!\!\!\!\!\!\!\!\!\!\!\!\!\!
\left[\,\rho_1^{\phantom{0}}(p) \,,\,{\cal H}_2\,\right]
\,=\,  \widetilde{g}_2 v_F^{\phantom{0}} \,p   \,\rho_2^{\phantom{0}}(p)
\;, \quad 
\left[\,\rho_2^{\phantom{0}}(-p) \,,\,{\cal H}_2\,\right]
\, = \,  \widetilde{g}_2 v_F^{\phantom{0}} \,p   \,\rho_1^{\phantom{0}}(-p)
\;,
\label{eq:commutator_2}
\rule{0cm}{0.5cm}
\end{eqnarray}
where
$\ \widetilde{g}_4 = g_4/(2\pi v_F^{\phantom{0}})\ $, and
$\ \widetilde{g}_2 = g_2/(2\pi v_F^{\phantom{0}})\ $.

\subsection{Two conservation laws}

The operator for the charge and current are defined by 
\begin{eqnarray}
\rho_{c}^{\phantom{0}}(p) \ = \ 
\rho_{1}^{\phantom{0}}(p) \,+\, \rho_{2}^{\phantom{0}}(p)
\;, \qquad \quad
\rho_{J}^{\phantom{0}}(p) \ = \ 
\rho_{1}^{\phantom{0}}(p) \, - \, \rho_{2}^{\phantom{0}}(p)
\;.
\label{eq:charge_current_TL}
\end{eqnarray}
In the real space, 
the operators for the left and right movers $\nu=1,\,2$ 
are written in the form 
\begin{eqnarray}
\rho_{\nu}^{\phantom{0}}(x) \, = \, 
{1 \over L} \sum_{p>0} \,
\Biggl(\,
\rho_{\nu}^{\phantom{0}}(p)\,e^{ipx}
\ + \ \rho_{\nu}^{\phantom{0}}(-p)\,e^{-ipx} \,\Biggr) \;.
\label{eq:real_space}
\end{eqnarray}
The equation of motion 
for $\rho_{c}^{\phantom{0}}(x)$ and $\rho_{J}^{\phantom{0}}(x)$ are 
derived from the Heisenberg equation using the commutation relations
in eqs.\ (\ref{eq:commutator_0})--(\ref{eq:commutator_2}), 
\begin{eqnarray}
{\partial \over \partial t}\, 
\rho_{c}^{\phantom{0}}(x,t) 
\, + \, 
v_J^{\phantom{0}}\,
{\partial \over \partial x}\, 
\rho_{J}^{\phantom{0}}(x,t) 
 \,=\, 0 \;,
&& \; 
v_J^{\phantom{0}} \, = \, 
v_F^{\phantom{0}} \left(\,1 + \widetilde{g}_4 - \widetilde{g}_2 \,\right) 
\;,
\label{eq:continuous_TL1}
\rule{1cm}{0cm}
\\
{\partial \over \partial t}\, 
\rho_{J}^{\phantom{0}}(x,t) 
 \, + \, v_N^{\phantom{0}}\,
{\partial \over \partial x}\, 
\rho_{c}^{\phantom{0}}(x,t) 
\,=\, 0 \;, && \; 
v_N^{\phantom{0}} \, = \, 
v_F^{\phantom{0}} \left(\,1 + \widetilde{g}_4 + \widetilde{g}_2 \,\right) 
\;.
\rule{0cm}{0.8cm}
\label{eq:continuous_TL2}
\end{eqnarray}
Because there are two independent equations for 
$\rho_{c}^{\phantom{0}}(x,t)$ and $\rho_{J}^{\phantom{0}}(x,t)$, 
the explicit form of these Heisenberg operators 
can be calculated analytically;
\begin{eqnarray}
&&
\!\!\!\!\!\!\!\!\!\!\!\!\!\!\!\!\!\!\!\!
\left(\, {\partial^2 \over \partial t^2}\, 
\,-\, v_{\rho}^{2} \, 
{\partial^2 \over \partial x^2 } \,\right)
\rho_{c}^{\phantom{0}}(x,t) \, = \, 0\,,
\qquad
\left(\, {\partial^2 \over \partial t^2}\, 
\,-\, v_{\rho}^{2} \, 
{\partial^2 \over \partial x^2 } \,\right)
\rho_{J}^{\phantom{0}}(x,t) \, = \, 0\,,
 \rule{0cm}{0.8cm}
\label{eq:wave_eq}
\\
&&
\!\!\!\!\!\!\!
v_{\rho}^{2} \, = \, v_J^{\phantom{0}} \,v_N^{\phantom{0}}
\;. \rule{0cm}{0.8cm}
\end{eqnarray}
The relation among the three velocities
 $v_{J}^{\phantom{0}}$, $v_{N}^{\phantom{0}}$ 
and $v_{\rho}^{\phantom{0}}$ can be summarized as
\begin{eqnarray}
v_{J}^{\phantom{0}} \, = \, 
K_{\rho}^{\phantom{0}} v_{\rho}^{\phantom{0}}\;,
\qquad 
v_{N}^{\phantom{0}} \, = \, 
{v_{\rho}^{\phantom{0}} \over K_{\rho}^{\phantom{0}} } \;,
\qquad  
K_{\rho}^{\phantom{0}} \, \equiv \, 
\sqrt{{ 
1 + \widetilde{g}_4 - \widetilde{g}_2 \over
1 + \widetilde{g}_4 + \widetilde{g}_2
}}\;\;.
\end{eqnarray}

\subsection{Charge and current correlation functions}

Owing to the property shown in eq.\ (\ref{eq:wave_eq}),
the correlation functions for the density operators 
\begin{eqnarray}
\chi_{\mu\nu}^r(p,t) \ = \
 i\, {1 \over L} 
\,\theta(t)
\left\langle \left[\,
\rho_{\mu}^{\phantom{0}}(p,t) \,,\,
\rho_{\nu}^{\phantom{0}}(-p) \,\right] \right\rangle 
\;, \qquad  \mbox{for}\ \ \mu,\nu\,=1,\,2 
\end{eqnarray}
can also be calculated exactly. The equation of motion 
for these correlations are given by
\begin{eqnarray}
& &
\!\!\!\!\!\!\!\!\!\!\!\!\!
i\,{\partial \over \partial t}\, 
\chi_{\mu\nu}^r(p,t) 
\nonumber \\
&& \ =\,   -\,{1 \over L} \,\delta(t)
\left\langle \left[\,
\rho_{\mu}^{\phantom{0}}(p) \,,\,
\rho_{\nu}^{\phantom{0}}(-p) \,\right] \right\rangle 
- 
\,\theta(t)
\,{1 \over L}
\left\langle \left[\,
{\partial \rho_{\mu}^{\phantom{0}}(p,t) \over \partial t}
\,,\,
\rho_{\nu}^{\phantom{0}}(-p) \,\right] \right\rangle ,
\nonumber
\\
 & & \ =\, - \,{p \over 2\pi}\, \tau^3_{\mu\nu} \, \delta(t)
\, +\,i
\,\theta(t)
\, {1 \over L} 
\left\langle \left[\,
\left[\rho_{\mu}^{\phantom{0}}(p,t)\,,\, {\cal H}_{\rm TL} \right]\,,\,
\rho_{\nu}^{\phantom{0}}(-p) \,\right] \right\rangle \;,
\label{eq:eqn_chi}
\rule{0cm}{0.8cm}
\end{eqnarray}
where $\mbox{\boldmath$\tau$}^3$ is a Pauli matrix:
\begin{equation}
\mbox{\boldmath$\tau$}^1 \, = \, 
 \left[ 
           \matrix { 0 & 1  \cr   
                     1 & 0 \cr } 
                                 \right] ,\ \ 
\mbox{\boldmath$\tau$}^2 \, = \, 
 \left[ 
           \matrix { 0 & -i  \cr   
                     i & \phantom{-}0 \cr } 
                                 \right] ,\ \ 
\mbox{\boldmath$\tau$}^3 \, = \, 
 \left[ 
           \matrix { 1 & \phantom{-}0  \cr   
                     0 & -1 \cr } 
                                 \right] ,\ \ 
 \mbox{\boldmath$1$} \, = \, 
 \left[ 
           \matrix { 1 & 0  \cr   
                     0 & 1 \cr } 
                                 \right] .
\end{equation}
The commutation relation 
$[ \rho_{\mu}^{\phantom{0}}(p,t)\,,\, {\cal H}_{\rm TL} ]$
in eq.\ (\ref{eq:eqn_chi}) can be calculated by using 
eqs.\ (\ref{eq:commutator_0})--(\ref{eq:commutator_2}).
Then, by carrying out the Fourier transform 
with respect to $t$, we obtain 
\begin{eqnarray}
& &
\!\!\!\!\!\!\!\!\!\!\!\!\!
\Biggl\{\ 
\omega\, 
\mbox{\boldmath$\tau$}^3 \ - \ 
v_{\rho}^{\phantom{0}} p 
\left(\,
 \cosh \varphi\, \mbox{\boldmath$1$}  
        \,+\, \sinh \varphi\, \mbox{\boldmath$\tau$}^1
        \, \right)
\ \Biggr\} \, 
\mbox{\boldmath$\chi$}^r(p,\omega) 
\ = \ -\,{p \over 2 \pi}\, 
\mbox{\boldmath$1$} \;,
\label{eq:eqn_chi_mat}
 \\
& & 
\cosh \varphi \equiv \ 
{1 + \widetilde{g}_4 \over 
\sqrt{(1 + \widetilde{g}_4)^2 - \widetilde{g}_2^2 } }
\ = \  {1\over 2}\left( {1 \over K_{\rho}}  \, + \, K_{\rho}  
        \right)
\;, 
\rule{0cm}{1cm}
\\
& &
\sinh \varphi \ \equiv \ 
{\widetilde{g}_2 \over 
\sqrt{(1 + \widetilde{g}_4)^2 - \widetilde{g}_2^2 } }
\ = \  {1\over 2}\left( {1 \over K_{\rho} } \, - \, K_{\rho}  
        \right)
\;.
\rule{0cm}{1cm}
\end{eqnarray}
Note that, 
$\cosh \varphi\, \mbox{\boldmath$1$}  
        \,+\, \sinh \varphi\, 
        \mbox{\boldmath$\tau$}^1 =
\exp\left( {\varphi\, \mbox{\boldmath$\tau$}^1} \right)$. 
The Bogoliubov transformation given by 
$\exp\left( {\varphi\, \mbox{\boldmath$\tau$}^1}/2 \right)$ 
has a property,
\begin{eqnarray}
\mbox{\boldmath$\tau$}^3 \,=\,
\exp\left( {{\varphi\, \mbox{\boldmath$\tau$}^1}\over 2} \right)
       \, \mbox{\boldmath$\tau$}^3 \,
\exp\left( {{\varphi\, \mbox{\boldmath$\tau$}^1} \over 2} \right)
\;.
\end{eqnarray}
Therefore, eq.\ (\ref{eq:eqn_chi_mat}) can be diagonalized, as
\begin{eqnarray}
& &
\!\!\!\!\!\!\!\!\!\!\!\!\!
\exp\left( {{\varphi\, \mbox{\boldmath$\tau$}^1} \over 2} \right)
\,
\left\{\,
\omega\, 
\mbox{\boldmath$\tau$}^3 \, - \, 
v_{\rho}^{\phantom{0}} p \, \mbox{\boldmath$1$}  
\, \right\} \, 
\exp\left( {{\varphi\, \mbox{\boldmath$\tau$}^1} \over 2} \right)
\,\mbox{\boldmath$\chi$}^r(p,\omega) 
\ = \ -\,{p \over 2 \pi}\, 
\mbox{\boldmath$1$} \;.
\label{eq:eqn_chi_mat2}
\end{eqnarray}
With this transformation by 
 $\exp\left( {\varphi\, \mbox{\boldmath$\tau$}^1}/2 \right)$, 
the operators $C_p^{\phantom{\dagger}}$ and $C_{-p}^{\dagger}$ 
are transformed into
\begin{eqnarray}
& &
\!\!\!\!\!\!\!\!\!\!
\left[ 
    \matrix { \gamma_p^{\phantom{\dagger}} \cr 
             \rule{0cm}{0.6cm} \gamma_{-p}^{\dagger}  \cr }   
 \right] 
\, = \, 
 \left[ \matrix { \cosh (\varphi/2) & \sinh (\varphi/2)  \cr   
                  \sinh (\varphi/2) & \cosh (\varphi/2)\rule{0cm}{0.8cm}\cr } 
 \right]\, 
 \left[ 
    \matrix { C_p^{\phantom{\dagger}} \cr 
             \rule{0cm}{0.6cm} C_{-p}^{\dagger}  \cr }   
 \right]  \;,
\label{eq:Bogoliubov}
\\
 & &
\!\!\!\!\!\!\!\!\!\!
\cosh (\varphi/2)
\, = \,  {1\over 2}\left( {1 \over \sqrt{K_{\rho}}}  \, + \, \sqrt{K_{\rho}}  
        \right) , \quad  \ \ 
\sinh (\varphi/2)
\, = \,  {1\over 2}\left( {1 \over \sqrt{K_{\rho}}}  \, - \, \sqrt{K_{\rho}}  
        \right) ,
\rule{0cm}{1cm}
\nonumber \\
\end{eqnarray}
where 
the bose statistics is preserved for the new operators, 
$\left[\gamma_p^{\phantom{\dagger}}\,,\, \gamma_{p'}^{\dagger} \right] 
\,=\, \delta_{pp'}$.
The explicit form of $\mbox{\boldmath$\chi$}^r(p,\omega)$ is 
determined by eq.\ (\ref{eq:eqn_chi_mat2}),
\begin{eqnarray}
&&
\!\!\!\!\!\!\!\!\!
\mbox{\boldmath$\chi$}^r(p,\omega) \,=\, 
-\,{p \over 2 \pi}\, 
\exp\left( -{{\varphi\, \mbox{\boldmath$\tau$}^1} \over 2} \right) 
\Biggl\{
D_{+}^r(p,\omega) \,
\mbox{\boldmath$\tau$}^3 
\, + \, 
D_{-}^r(p,\omega)\, 
 \mbox{\boldmath$1$}  
 \Biggr\} 
\exp\left( -{{\varphi\, \mbox{\boldmath$\tau$}^1} \over 2} \right) 
\nonumber
\\
&& \qquad \ =\,
-\,{p \over 2 \pi}\, 
\Biggl\{\,
D_{+}^r(p,\omega) \,
\mbox{\boldmath$\tau$}^3 
\, + \, 
D_{-}^r(p,\omega)\, 
\left(\, \cosh \varphi\, \mbox{\boldmath$1$}  
        \,-\, \sinh \varphi\, \mbox{\boldmath$\tau$}^1\,\right)
\, \Biggr\} \;, 
\nonumber
\rule{0cm}{0.9cm}
\\
&& 
\label{eq:eqn_chi_mat3}
\\
&&
\!\!\!\!\!\!\!\!\!
 D_{\pm}^r(p,\omega) \,=\, {1 \over 2} \left(\,
 {1 \over \omega - v_{\rho}^{\phantom{0}} p + i\delta} \ \pm \  
 {1 \over \omega + v_{\rho}^{\phantom{0}} p + i\delta} \,\right) \;.
\end{eqnarray}
The charge susceptibility $\chi_c^r(p,\omega)$,
which corresponds to the 
$\rho_{c}^{\phantom{0}}$-$\rho_{c}^{\phantom{0}}$ correlation function,
is given by 
\begin{equation}
\chi_c^r(p,\omega) \ = \ \sum_{\mu\nu} \chi_{\mu\nu}^r(p,\omega)
\ = \  -\,{K_{\rho}\over \pi v_{\rho}^{\phantom{0}} }\ 
{(v_{\rho}^{\phantom{0}} p )^2 
 \over 
 (\omega + i\delta)^2\, - \, ( v_{\rho}^{\phantom{0}} p )^2} 
 \;.
\end{equation}
Then, the uniform charge susceptibility is given by   
${\displaystyle \lim_{p\to 0}}\, 
\chi_c^r(p,0)\,=\,K_{\rho}/ (\pi v_{\rho}^{\phantom{0}})$.
It becomes twice as large for the spin $1/2$ fermions.

The current operator is determined by eqs.\ 
(\ref{eq:charge_current_TL})--(\ref{eq:continuous_TL1}) as
\begin{equation}
J \,=\, e\,v_{J}^{\phantom{0}}\rho_{J}^{\phantom{0}}\;.
\end{equation}
Therefore, the $J$-$J$ correlation function is given by 
\begin{equation}
K^r(p,\omega) \ = \ 
 -\,{e^2 K_{\rho}v_{\rho}^{\phantom{0}}\over \pi}\ 
{(v_{\rho}^{\phantom{0}} p )^2 
 \over 
 (\omega + i\delta)^2\, - \, ( v_{\rho}^{\phantom{0}} p )^2} 
 \;.
\end{equation}
Then, the conductivity can be calculated with the Kubo formula,
\begin{eqnarray}
\sigma(p,\omega) 
&=& 
{K^r(p,\omega) - K^r(p,0) \over i\omega}
\nonumber \\
&=& 
  {e^2 K_{\rho}v_{\rho}^{\phantom{0}} \over \pi}\, 
{i\,\omega 
 \over 
 (\omega + i\delta)^2\, - \, ( v_{\rho}^{\phantom{0}} p )^2
 } 
 \;. \rule{0cm}{0.8cm} 
\end{eqnarray}
The conductivity $\sigma(p,\omega)$ for 
 a uniform $p=0$ and stationary $\omega=0$ field  
depends on the order of taking 
the limits of  $p \to 0$ and $\omega \to 0$. 
The Drude weight corresponds to the $p\to 0$ limit,
\begin{equation}
\mbox{Re}\,\sigma(0,\omega) \,=\,   
e^2 K_{\rho} v_{\rho}^{\phantom{0}} \, \delta(\omega)\;,
\end{equation}
In the real space, the conductivity takes the form 
\begin{eqnarray}
\sigma(x,\omega) 
&=& 
\int_{-\infty}^{\infty} {dp \over 2 \pi}\ \sigma(p,\omega) \,e^{ipx} 
\nonumber
\\
&=& 
\int_{-\infty}^{\infty}{dp \over 2 \pi}\,
 {i\,e^2 K_{\rho}v_{\rho}^{\phantom{0}}\over 2\pi}\ 
\left[\,
{1  \over  \omega  \,- \,  v_{\rho}^{\phantom{0}} p + i\delta} 
\,+\,
{1  \over  \omega  \,+ \,  v_{\rho}^{\phantom{0}} p + i\delta} 
\,\right]
\,e^{ipx}
\nonumber
\\
&=& 
 {e^2 K_{\rho}\over 2\pi}\, 
 e^{i {\omega \over v_{\rho}} |x|} \;\;.
\rule{0cm}{0.6cm} 
\end{eqnarray}
The dc conductance corresponds to the $\omega \to 0$ limit,
\begin{eqnarray}
\sigma(x,0) 
\ = \  
 {e^2\over 2\pi \hbar}\, K_{\rho}
  \ = \ {e^2\over h}\, K_{\rho}\;,
\end{eqnarray}
where $\hbar$ has been reinserted.

\subsection{Boson representation of the Hamiltonian}

We have seen in the above  
that the bosonic excitations play an important role 
on the transport properties of the Tomonaga-Luttinger model. 
Correspondingly, there is one notable feature 
in the commutation relations for the density operators 
in eqs.\ (\ref{eq:commutator_0}) and (\ref{eq:commutator_4}):  
the two parts of the Hamiltonian 
 ${\cal H}_0$ and ${\cal H}_4 / \widetilde{g}_4$ show the same 
commutation relations.
Therefore,  one can introduce 
an effective Hamiltonian $\widetilde{\cal H}_0$ defined by 
\begin{eqnarray}
\widetilde{\cal H}_0
\ = \ 
{2 \pi v_F^{\phantom{0}} \over L}\, 
\sum_{p>0}\,
\rho_1^{\phantom{0}}(-p)\, 
\rho_1^{\phantom{0}}(p) 
\ + \ 
{2 \pi v_F^{\phantom{0}} \over L}\, 
\sum_{p>0}\,
\rho_2^{\phantom{0}}(-p)\, 
\rho_2^{\phantom{0}}(p) \;,
\label{eq:H0_TL_boson}
\end{eqnarray}
which reproduces the commutation relation eq.\ (\ref{eq:commutator_0}).
Thus, the correlation functions 
can be calculated exactly by
using $\widetilde{\cal H}_0$ as a replacement for ${\cal H}_0$.
The effective Hamiltonian is written 
in a bilinear form with the boson operators,
\begin{eqnarray}
\widetilde{\cal H}_{\rm TL} &\equiv& 
\widetilde{\cal H}_{0} \,+\, {\cal H}_{4} \,+\, {\cal H}_{2} 
\nonumber
\\
&=& 
\sum_{p>0}
 \left[ \!\! 
 \begin{array}[t]{l}
       C_p^{\dagger} \quad C_{-p}^{\phantom{\dagger}} 
 \end{array}
 \!\! \right]  \,
v_{\rho}^{\phantom{0}} p 
\left(\, \cosh \varphi\, \mbox{\boldmath$1$}  
        \,+\, \sinh \varphi\, \mbox{\boldmath$\tau$}^1\,\right)\,
%
 \left[\! 
 \begin{array}{l}
   C_p^{\phantom{\dagger}} \\
   C_{-p}^{\dagger}      
 \end{array}
 \!\right]  
 \nonumber
 \\
 &=& 
\sum_{p>0}
v_{\rho}^{\phantom{0}} p  
\left(\,  \gamma_p^{\dagger}\,\gamma_p^{\phantom{\dagger}}  
          \, + \, 
          \gamma_{-p}^{\dagger}\,\gamma_{-p}^{\phantom{\dagger}} 
 \,\right)  \ + \; \mbox{const}   \quad .
\end{eqnarray}

In this section, we have discussed only 
the two-particle correlation functions.
The equation of motion for 
the single-particle Green's function can also be written 
in a closed form \cite{GL,ES}, and  
the precise calculations have been reported in Refs.\ \cite{Voit,Meden}.
Furthermore, 
for comprehensive description of the bosonization, 
see Ref.\ \cite{Haldane}.

\end{document}